\documentclass[twolcolumn,apj]{emulateapj}
\usepackage{ctable}
\usepackage{amsmath}
\usepackage{graphicx}

\pdfoutput=1

\begin{document}

\title{What Next-Generation 21~cm Power Spectrum Measurements Can Teach Us
About the Epoch of Reionization}

\author{Jonathan C. Pober\altaffilmark{1,15},
Adrian Liu\altaffilmark{2,3}, 
Joshua S. Dillon\altaffilmark{4,5},
James E. Aguirre\altaffilmark{6},
Judd D. Bowman\altaffilmark{7},
Richard F. Bradley\altaffilmark{8,9,10},
Chris L. Carilli\altaffilmark{11,12},
David R. DeBoer\altaffilmark{13},
Jacqueline N. Hewitt\altaffilmark{4,5},
Daniel C. Jacobs\altaffilmark{7},
Matthew McQuinn\altaffilmark{2,16},
Miguel F. Morales\altaffilmark{1},
Aaron R. Parsons\altaffilmark{2},
Max Tegmark\altaffilmark{4,5},
Dan J. Werthimer\altaffilmark{14}
}
\altaffiltext{1}{Physics Department, University of Washington, Seattle, WA}
\altaffiltext{2}{Department of Astronomy, University of California Berkeley, Berkeley, CA}
\altaffiltext{3}{Berkeley Center for Cosmological Physics, University of California Berkeley, Berkeley, CA}
\altaffiltext{4}{Department of Physics, Massachusetts Institute of Technology, Cambridge, MA}
\altaffiltext{5}{MIT Kavli Institute, Massachusetts Institute of Technology, Cambridge, MA}
\altaffiltext{6}{Department of Physics and Astronomy, University of Pennsylvania, Philadelphia, PA}
\altaffiltext{7}{School of Earth and Space Exploration, Arizona State University, Tempe, AZ}
\altaffiltext{8}{Astronomy Department, University of Virginia, Charlottesville, VA}
\altaffiltext{9}{Department of Electrical and Computer Engineering, University of Virginia, Charlottesville, VA}
\altaffiltext{10}{National Radio Astronomy Observatory, Charlottesville, VA}
\altaffiltext{11}{National Radio Astronomy Observatory, Socorro, NM}
\altaffiltext{12}{Astrophysics Group, Cavendish Laboratory, University of Cambridge, Cambridge, UK}
\altaffiltext{13}{Radio Astronomy Laboratory, University of California Berkeley, Berkeley, CA}
\altaffiltext{14}{Space Sciences Laboratory, University of California Berkeley, Berkeley, CA}

\altaffiltext{15}{National Science Foundation Astronomy and Astrophysics
Postdoctoral Fellow}
\altaffiltext{16}{Hubble Fellow}

\begin{abstract}
A number of experiments are currently working towards a measurement of the 21~cm signal
from the Epoch of Reionization.  Whether or not
these experiments deliver a detection of cosmological emission, their
limited sensitivity will prevent them from providing detailed
information about the astrophysics of reionization.  In this work, we consider
what types of measurements will be enabled by a next-generation of larger
21~cm EoR telescopes.  To calculate the type of constraints that will
be possible with such arrays, we use simple models for the instrument,
foreground emission, and the reionization history.
We focus primarily on an instrument modeled after the $\sim 0.1~\rm{km}^2$ collecting
area Hydrogen
Epoch of Reionization Array (HERA) concept design,
and parameterize the uncertainties with regard to foreground emission
by considering different limits to the recently described ``wedge"
footprint in $k$-space.
Uncertainties in the reionization history are accounted for using a series
of simulations which vary the ionizing efficiency and minimum virial 
temperature of the galaxies responsible for reionization,
as well as the mean free path of ionizing photons through the IGM.
Given various combinations of models, we consider the significance of the
possible power spectrum detections, the ability to trace the power spectrum
evolution versus redshift, the detectability of salient power spectrum
features, and the achievable level of quantitative constraints on astrophysical 
parameters.  Ultimately, we find that $0.1~\rm{km}^2$ 
of collecting area is enough to ensure
a very high significance ($\gtrsim30\sigma$) detection
of the reionization power spectrum in even the most pessimistic scenarios.
This sensitivity should
allow for meaningful constraints on the reionization history and astrophysical
parameters, especially if foreground subtraction techniques can be improved
and successfully implemented.\\
\end{abstract}


\section{Introduction}

The Epoch of Reionization (EoR) represents a turning point in cosmic history,
signaling the moment when large scale structure has become significant
enough to impart a global change to the state of the baryonic universe.  In particular,
the EoR is the period when ultraviolet photons (likely from the first galaxies)
reionize the neutral hydrogen in the intergalactic medium (IGM).
As such, measurements
of the conditions during the EoR promise a wealth of information about
the evolution of structure in the universe. 
Observationally, the redshift of EoR is roughly constrained to be between
$z \sim 6 \mbox{--} 13$, with a likely extended duration; see 
\cite{furlanetto_et_al_2006b}, \cite{barkana_and_loeb_2007},
and \cite{loeb_and_furlanetto_2013} for 
reviews of the field.
Given the difficulties of optical/NIR observing at these
redshifts, the highly-redshifted 21~cm line of neutral hydrogen has been
recognized as a unique probe of the conditions during the EoR 
(see \citealt{morales_and_wyithe_2011} and 
\citealt{pritchard_and_loeb_2012} for recent reviews discussing this technique).

In the last few years, the first generation of experiments targeting
a detection of this highly-redshifted 21~cm signal from the EoR 
has come to fruition.  In particular, the LOw Frequency ARray 
(LOFAR; \citealt{yatawatta_et_al_2013,van_haarlem_et_al_2013})\footnote{http://www.lofar.org/}, 
the Murchison Widefield Array (MWA; \citealt{tingay_et_al_2013,bowman_et_al_2013})\footnote{http://www.mwatelescope.org/}, 
and the Donald C. Backer Precision Array for Probing the Epoch of Reionization
(PAPER; \citealt{parsons_et_al_2010})\footnote{http://eor.berkeley.edu/}
have all begun long, dedicated campaigns with the goal of detecting the
21~cm power spectrum.  Ultimately, the success or failure of these
campaigns will depend on the feasibility of controlling both instrumental
systematics and foreground emission.
But even if these challenges can be overcome, a positive detection of
the power spectrum will likely be marginal at best because of limited
collecting area.  Progressing from a detection to a characterization of the
power spectrum (and eventually, to the imaging of the EoR) will require
a next generation of larger 21~cm experiments.

The goal of this paper is to explore the range of constraints that
could be achievable with larger 21~cm experiments and, in particular, focus
on how those constraints translate into a physical understanding of the EoR.  
Many groups have
analyzed the observable effects of different reionization models on the 21~cm
power spectrum; see e.g., 
\citet{zaldarriaga_et_al_2004}, \citet{furlanetto_et_al_2004},
\citet{mcquinn_et_al_2006},
\citet{bowman_et_al_2006}, \citet{bowman_et_al_2007}, 
\citet{trac_and_cen_2007}, \citet{lidz_et_al_2008}, and
\citet{iliev_et_al_2012}.  These studies did not include the
more sophisticated understanding of foreground emission that has arisen
in the last few years, i.e., the division of 2D cylindrical $k$-space into the
foreground-contaminated ``wedge" and the relatively clean ``EoR window"
\citep{datta_et_al_2010,vedantham_et_al_2012,morales_et_al_2012,parsons_et_al_2012b,trott_et_al_2012,thyagarajan_et_al_2013}.
The principal undertaking of this present work is to reconcile
these two literatures, exploring the effects of both different 
EoR histories and foreground removal models on the recovery of astrophysical
information from the 21~cm power spectrum.
Furthermore, in this work we present some of the first analysis focused 
on using realistic measurements to distinguish between different theoretical 
scenarios, rather than simply computing observable (but possibly degenerate) 
quantities from a given theory.
The end result is a set of generic
conclusions that both demonstrates the need for a large collecting area next
generation experiment and motivates the 
continued development of foreground removal algorithms.

In order to accomplish these goals, this paper will employ simple models
designed to encompass a wide range of possible scenarios.  These models
are described in \S\ref{sec:models}, wherein we describe the
models for the
instrument (\S\ref{sec:instrument}), foregrounds (\S\ref{sec:foregrounds}),
and reionization history (\S\ref{sec:eor}).  In \S\ref{sec:results}, we
present a synthesis of these models and the resultant range of potential
power spectrum constraints, including a detailed examination
of how well one can recover physical parameters describing the EoR in 
\S\ref{sec:adrian}.
In \S\ref{sec:conclusions}, we conclude with several generic messages
about the kind of science the community can expect from 21~cm experiments
in the next $\sim5$ years.

\section{The Models}
\label{sec:models}

In this section we present the various models for the 
instrument (\S\ref{sec:instrument}), foreground removal 
(\S\ref{sec:foregrounds}),
and reionization history (\S\ref{sec:eor}) used to explore
the range of potential EoR measurements.  
In general, these models are chosen not because they
necessarily mirror
specific measurements or scenarios, but rather because of their
simplicity while still encompassing a wide range of uncertainty about many 
parameters.
We choose several different parameterizations of the foreground removal
algorithms, and use simple simulations to probe a wide variety of reionization
histories.
Our model telescope (described below in \S\ref{sec:instrument}) 
is based off the proposed Hydrogen Epoch of Reionization Array (HERA);
we present sensitivity calculations and astrophysical constraints for
other 21~cm experiments in the appendix.

\subsection{The Telescope Model}
\label{sec:instrument}

\begin{figure}[ht!]
\centering
\includegraphics[width=3.5in,trim=1cm 0cm 1cm 0cm,clip=True]{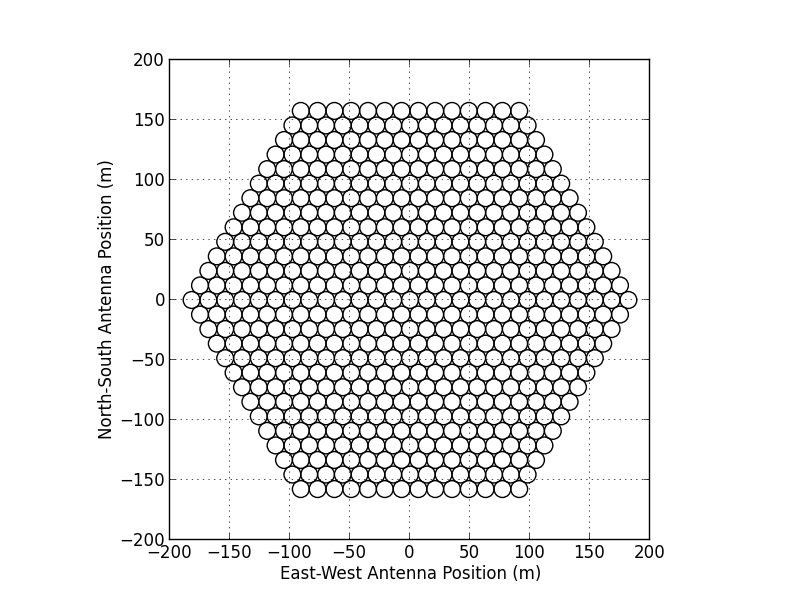}
\caption{The 547-element, hexagonally packed HERA concept design, with 14~m
reflector elements.  Outrigger antennas may be included in the final design
for the purposes of foreground imaging, but they are not treated here, since
they add little to power spectrum sensitivity.}
\label{fig:hex547}
\end{figure}

The most significant difference between the current and next generations 
of 21~cm instruments
will be a substantial increase in collecting area and, therefore, sensitivity. 
In the main body of this work, we use an instrument modeled after a concept design for the
Hydrogen Epoch of Reionization Array (HERA)\footnote{http://reionization.org/}.
This array consists of 547 zenith-pointing
14~m diameter reflecting-parabolic elements in a close-packed hexagon, as shown in Figure
\ref{fig:hex547}. 
The total collecting area of this array is $84,000~\rm{m}^2$, or approximately
one tenth of a square kilometer.  The goal of this work is not to justify
this particular design choice, but rather to show that this scale instrument
enables a next level of EoR science beyond the first generation experiments.
In the appendix, we present the resultant sensitivities and achievable
constraints on the astrophysical parameters of interest for
several other 21~cm telescopes: PAPER, the MWA, LOFAR, and a concept
design for the SKA-Low Phase 1.  Generically, we find that
power spectrum sensitivities are a strong function of array configuration,
especially compactness and redundancy.  However, once the
power spectrum sensitivity of an array is known, constraints
on reionization physics appear to be roughly independent of other
paramters.

In many ways, the HERA concept array design
is quite representative of 21~cm EoR
experiments over the next $\sim5\mbox{--}10$ years.  As mentioned, it has
a collecting area of order a tenth of a square kilometer --- 
significantly larger than any current instrument,
but smaller than Phase 1 of the low-frequency Square
Kilometre Array (SKA1-low)\footnote{http://www.skatelescope.org/}.
(See Table \ref{tab:instruments} for a summary of different
EoR telescopes.)
In terms of power spectrum sensitivity,
\cite{parsons_et_al_2012a} demonstrated the power of array
redundancy for reducing thermal noise uncertainty, and showed
that a hexagonal configuration has the greatest instantaneous redundancy.
In this sense, the HERA concept design is optimized for power spectrum
measurements.
Other configurations in the literature have been optimized for foreground
imaging or other additional science; the purpose of this work is not to 
argue for or against these designs.  
Rather, we concentrate primarily on science with the 21~cm power spectrum, and
use the HERA concept design as representative of power spectrum-focused
experiments.
Obviously,
arrays with more (less) collecting area will have correspondingly greater
(poorer) sensitivity.  
The key parameters of our fiducial concept array
are given in Table \ref{tab:array}, and constraints achievable
with other arrays are presented in the appendix.

\ctable[caption=Fiducial System Parameters, label=tab:array, pos=ht, mincapwidth=3.5in]{c|c}%
{}
{\hline
Observing Frequency & $50 \mbox{--} 225$ MHz \\
$T_{\rm receiver}$ & 100 K \\
Parabolic Element Size & 14~m \\
Number of Elements & 547 \\
Primary beam size & $8.7^\circ$ FWHM at 150~MHz \\
Configuration & Close-packed hexagonal \\
Observing mode & Drift-scanning at zenith \\
\hline}

\subsubsection{Calculating Power Spectrum Sensitivity}
\label{sec:sense_calc}

To calculate the power spectrum sensitivity of our fiducial array, we use
the method presented in \cite{pober_et_al_2013a}, which is briefly summarized
here.  This method begins by creating the $uv$ coverage of the observation
by gridding each baseline into the $uv$ plane,
including the effects of earth-rotation synthesis over the course
of the observation.  
We choose $uv$ pixels the size of the antenna element in wavelengths,
and assume that any baseline samples only one pixel at a time.
Each pixel is treated as an
independent sample of one $k_{\perp}$-mode, along which the instrument
samples a wide range of $k_{\parallel}$-modes 
specified by the observing bandwidth.
The sensitivity to any one mode of the dimensionless power spectrum 
is given by Equation (4) in \cite{pober_et_al_2013a}, which is in turn
derived from Equation (16) of \cite{parsons_et_al_2012a}:
\begin{equation}
\label{eq:sensitivity}
\Delta^2_{\rm N}(k) \approx X^2Y\frac{k^3}{2\pi^2}\frac{\Omega'}{2t}T_{\rm sys}^2,
\end{equation}
where $X^2Y$ is a cosmological scalar converting observed bandwidths and solid
angles to $h{\rm Mpc}^{-1}$, $\Omega ' \equiv \Omega_{\rm p}^2/\Omega_{\rm pp}$
is the solid angle of the power primary beam ($\Omega_{\rm p}$) squared, divided by the solid
angle of the square of the power primary beam ($\Omega_{\rm pp}$),\footnote{Although \cite{parsons_et_al_2012a}
and \cite{pober_et_al_2013a}
originally derived this relation with the standard power primary beam
$\Omega$, it was shown in \cite{parsons_et_al_2013} that the power-squared
beam enters into the correct normalizing factor.}
$t$ is the integration time on that particular $k$-mode, and $T_{\rm sys}$ is
the system temperature.  It should also be noted that this equation
is dual-polarization, i.e., it assumes both linear polarizations are measured
simultaneously and then combined to make a power spectrum estimate. 
Similar forms of this equation appear in \cite{morales_2005}
and \cite{mcquinn_et_al_2006}
which differ only by the polarization factor and power-squared primary
beam correction. 

In our formalism, each measured mode is attributed a noise value
calculated from Equation \ref{eq:sensitivity} (see 
\S\ref{sec:observation}
for specifics on the values of each parameter). 
Independent modes can be combined in quadrature to form spherical or
cylindrical power spectra as desired.  
One has a choice of how to combine non-instantaneously 
redundant baselines which do in fact sample the
same $k_{\perp}/uv$ pixel.  Such a situation can arise either through
the effect of the gridding kernel creating overlapping $uv$ footprints on
similar length baselines (``partial coherence"; 
\citealt{hazelton_et_al_2013}), or through the effect of earth-rotation
bringing a baseline into a $uv$ pixel previously sampled by another baseline.
Na\"{i}vely, this formalism treats these samples as perfectly coherent, i.e.,
we add the integration time of each baseline within a $uv$ pixel.  As suggested
by \cite{hazelton_et_al_2013}, however, it is possible that
this kind of simple treatment could
lead to foreground contamination in a large number of Fourier modes.
To explore the ramifications of this effect, 
we will also consider a case where only baselines which are 
instantaneously redundant are added coherently, and all other measurements
are added in quadrature when binning.  We discuss this model more in 
\S\ref{sec:foregrounds}.  

Since this method of calculating power spectrum sensitivities naturally
tracks the number of independent modes measured, sample variance is easily
included when combining modes by adding the value
of the cosmological power spectrum to each $(u,v,\eta)$-voxel
(where $\eta$ is the line-of-sight Fourier mode)
before doing any binning.  (Note that in the case where
only instantaneously redundant baselines are added coherently, partially
coherent baselines do not count as independent samples for the purpose of 
calculating sample variance.)   
Unlike \cite{pober_et_al_2013a}, we do not include the
effects of redshift-space distortions in boosting the line of sight
signal, since they will not boost the power spectrum of ionization 
fluctuations,
which is likely to dominate the 21~cm power spectrum at these redshifts.
We also ignore other second order effects on the power spectrum,
such as the ``light-cone" effect
\citep{datta_et_al_2012,la_plante_et_al_2013}.

\subsubsection{Telescope and Observational Parameters}
\label{sec:observation}

For the instrument value of $T_{\rm sys}$ we sum a frequency independent
100~K receiver temperature with a frequency dependent sky temperature,
$T_{\rm sky} = 60\rm{K}~(\lambda/1~{\rm m})^{2.55}$
\citep{thompson_et_al_2007}, giving a sky temperature of
351~K at 150~MHz.  Although this model is
$\sim100$~K lower than the system measured by \cite{parsons_et_al_2013},
it is consistent with recent LOFAR measurements \citep{yatawatta_et_al_2013,van_haarlem_et_al_2013}.  Since the smaller field of view of HERA will lead
to better isolation of a Galactic cold patch, we choose this
empirical relation for our model.

For the primary beam, we use a simple Gaussian model with a Full-Width Half-Max
(FWHM) of $1.06\lambda/D = 8.7^{\circ}$ at 150 MHz.  We assume the beam
linearly evolves in shape as a function of frequency.  In the actual
HERA instrument design, the PAPER dipole serves as a feed to the larger parabolic
element.  Computational E\&M modeling suggests this setup will have a beam
with FWHM of $9.8^{\circ}$.  Furthermore, the PAPER dipole response is 
specifically designed to evolve more slowly with frequency than our linear
model.  Although the frequency dependence of the primary beam enters into
our sensitivity calculations in several places (including the pixel size in
the $uv$ plane), the dominant effect is to change the normalization of the
noise level in Equation \ref{eq:sensitivity}.  For an extreme case with
no frequency evolution in the primary beam size (relative to 150~MHz), 
we find that the resultant
sensitivities increase by up to 40\% at 100 MHz (due to a
smaller primary beam than the linear evolution model), and decrease by up
to 30\% at 200 MHz (due to larger beam).  While all
instruments will have some degree of primary beam evolution as a function
of frequency, this extreme model demonstrates that some of the poor
low-frequency (high-redshift) sensitivities reported below can be partially 
mitigated by a more frequency-independent instrument design (although at the
expense of sensitivity at higher frequencies).

It should be pointed out that for snap-shot observations,
the large-sized HERA dishes prevent
measurements of the largest transverse scales.  
At 150 MHz ($z = 8.5$), the minimum
baseline length of 14~m corresponds to a transverse $k$-mode of
$k_{\perp} = 0.0068 h\rm{Mpc}^{-1}$.  
This array will be unable to observe transverse modes on larger scales,
without mosaicing or otherwise integrating over longer
than one drift through the primary beam.
The sensitivity calculation used in this work does not account for
such an analysis, and therefore will limit the sensitivity
of the array to larger-scale modes.
For an experiment targeting unique cosmological information on the largest
cosmic scales (e.g. primordial non-Gaussianity), this effect may prove
problematic.  For studies of the EoR power spectrum, the
limitation on measurements at low $k_{\perp}$ has little effect on the end result, especially given
the near ubiquitous presence of
foreground contamination on large-scales in our models
(\S\ref{sec:foregrounds}).

The integration time $t$ on a given $k$ mode,
is determined by the length of time any baseline in the array samples each
$uv$ pixel over the course of the observation.  Since we assume a drift-scanning
telescope, the length of the observation is set by the
size of the primary beam.  The time
it takes a patch of sky to drift through the beam is the duration over which
we can average coherently.  For the $\sim 10^{\circ}$ primary beam model above, 
this time is $\sim40$ minutes.

We assume that there exists one Galactic ``cold patch" spanning 6 hours in 
right ascension suitable for EoR observations, an assumption
which is based on measurements from both PAPER and the MWA and on previous
models (e.g. \citealt{de_oliveira-costa_et_al_2008}).  There are thus 9
independent fields of 40 minutes in right ascension (corresponding
to the primary beam size calculated above) 
which are observed per day.  We also assume EoR-quality
observations can only be conducted at night, yielding $\sim180$ days per year
of good observing.  Therefore, our thermal noise uncertainty 
(i.e. the $1\sigma$ error bar on the power spectrum) is reduced
by a factor of $\sqrt{9}\times 180$ over that calculated from one field,
whereas the contribution to the errors from sample variance
is only reduced by $\sqrt{9}$.

\subsection{Foregrounds}
\label{sec:foregrounds}

\begin{figure}[ht!]
\centering
\includegraphics[width=3.25in]{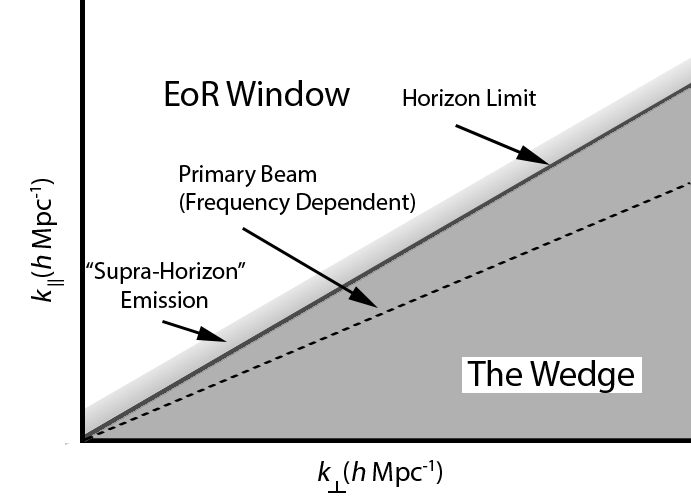}
\caption{A schematic diagram of the wedge and EoR window in 2D $k$-space.
See \S\ref{sec:foregrounds} for explanations of the terms.}
\label{fig:wedge}
\end{figure}

Because of its spectral smoothness, foreground emission is expected to
contaminate low order line-of-sight Fourier modes in the power spectrum.
Of great concern, though, are chromatic effects in an interferometer's
response, which can introduce spectral structure into foreground emission.
However, recent work has shown that these chromatic mode-mixing effects do not
indiscriminately corrupt all the modes of the power spectrum. 
Rather, 
foregrounds are confined to a ``wedge"-shaped region in the 
2D $(k_{\perp},k_{\parallel})$ plane, with more $k_{\parallel}$ modes free
from foreground contamination on the shortest baselines (i.e. at the smallest
$k_{\perp}$ values)
 \citep{datta_et_al_2010,vedantham_et_al_2012,morales_et_al_2012,parsons_et_al_2012b,trott_et_al_2012}, 
as schematically diagrammed in Figure \ref{fig:wedge}.
Power spectrum analysis in both \cite{dillon_et_al_2013b} and
\cite{pober_et_al_2013b} reveal the presence of the wedge in actual 
observations.  The single-baseline approach 
\citep{parsons_et_al_2012b} used in \cite{pober_et_al_2013b} yields a cleaner
EoR window, although at the loss of some sensitivity that comes from
combining non-redundant baselines.

However, there is still considerable debate about where to define the ``edge"
of the wedge.  Our three foreground models --- summarized in Table 
\ref{tab:fg}
--- differ in their choice of ``wedge edge."  Our pessimistic model
also explores the possibility that systematic effects discussed in
\cite{hazelton_et_al_2013} could prevent to coherent addition of
partially redundant baselines.  
It should be noted that although we use the shorthand ``foreground model" to
describe these three scenarios, in many ways these represent 
\emph{foreground removal models}, since they pertain to improvements
over current analysis techniques that may better separate foreground emission
from the 21~cm signal.

\subsubsection{Foreground Removal Models}

At present, 
observational limits on the ``edge" to the foreground wedge
in cylindrical $(k_{\perp},k_{\parallel})$-space are still somewhat
unclear.
\cite{pober_et_al_2013b} find the wedge to extend as much
as $\Delta k_{\parallel} = 0.05 \mbox{--} 0.1~h{\rm Mpc}^{-1}$
beyond the ``horizon limit," i.e., the $k_{\parallel}$ mode on a given baseline
that corresponds to the chromatic sine wave created by a 
flat-spectrum source of emission
located at the horizon.  (This mode in many ways represents a fundamental
limit, as the interference pattern cannot oscillate any faster for a 
flat-spectrum source of celestial emission; see \citealt{parsons_et_al_2012b}
for a full discussion of the wedge in the language of geometric delay space.)
Mathematically, the horizon limit is:
\begin{equation}
k_{\parallel,\rm{hor}} = \frac{2\pi}{Y}\frac{|\vec{b}|}{c} = \left(\frac{1}{\nu}\frac{X}{Y}\right) k_{\perp},
\end{equation}
where $|\vec{b}|$ is the baseline length in meters, $c$ is the speed of light,
$\nu$ is observing frequency, and $X$ and $Y$ are the previously described
cosmological scalars for converting observed bandwidths and solid
angles to $h{\rm Mpc}^{-1}$, respectively, defined in 
\cite{parsons_et_al_2012a} and \cite{furlanetto_et_al_2006b}.
\cite{pober_et_al_2013b} attribute the presence of ``supra-horizon" emission 
--- emission at $k_{\parallel}$ values greater than the horizon limit --- 
to spectral structure in the
foregrounds themselves, which creates a convolving kernel in $k$-space.
\cite{parsons_et_al_2012b} predict that the wedge could extend as much as 
$\Delta k_{\parallel} = 0.15~h{\rm Mpc}^{-1}$ beyond the horizon limit at the level of the 21~cm
EoR signal.  This supra-horizon emission has a dramatic effect on the size
of the EoR window, increasing the $k_{\parallel}$ extent of the wedge
by nearly a factor of 4 on the 
$16\lambda$-baselines used by PAPER in \cite{parsons_et_al_2013}.

Others have argued that the wedge will extend not to the horizon limit,
but only to the edges of the field-of-view, outside of which emission is
too attenuated to corrupt the 21~cm signal.  If achievable, 
this smaller wedge has a dramatic effect on sensitivity, since
theoretical considerations suggest that signal-to-noise
decreases quickly with increasing $k_{\perp}$ and $k_{\parallel}$. 
If one compares
the sensitivity predictions in \cite{parsons_et_al_2012b} for PAPER-132 and 
\cite{beardsley_et_al_2013} for MWA-128 (two comparably sized arrays), one
finds that these two different wedge definitions account for a large portion
of the difference between a marginal $2\sigma$ EoR 
detection and a 14$\sigma$ one.

While clearly inconsistent with the current results in
\cite{pober_et_al_2013b}, such a small wedge may be achievable
with new advances in foreground subtraction.
A large literature of work has gone into studying the removal of foreground
emission from 21~cm data (e.g. \citealt{morales_et_al_2006}, 
\citealt{bowman_et_al_2009}, \citealt{liu_et_al_2009}, 
\citealt{liu_and_tegmark_2011}, \citealt{chapman_et_al_2012},
\citealt{dillon_et_al_2013a}, \citealt{chapman_et_al_2013}).  
If successful, these techniques offer the
promise of working within the wedge.  However, despite the huge sensitivity boost,
working within the wedge clearly presents additional challenges beyond 
simply working
within the EoR window.  Working within the EoR window requires only keeping
foreground leakage from within the wedge to a level below the 21~cm signal;
the calibration challenge for this task can be significantly reduced by
techniques which are allowed to remove EoR signal from within the wedge
\citep{parsons_et_al_2013}.  Working within the wedge requires
foreground removal with up to 1 part in $10^{10}$
accuracy (in mK$^2$) while leaving the 21~cm signal unaffected.  Ensuring
that calibration errors do not introduce covariance between modes is
potentially an even more difficult task.  Therefore, 
given the additional effort it will take to be convinced that a residual 
excess of power post-foreground subtraction can be attributed to the EoR,
it seems
plausible that the first robust detection and measurements of the 21~cm EoR
signal will come from modes outside the wedge.  

To further complicate the issue, several effects have been identified
which can scatter power from the wedge into the EoR window. 
\cite{moore_et_al_2013} 
demonstrate how combining redundant visibilities without image plane correction (as done by PAPER) can corrupt the EoR signal outside the wedge, due to the
effects of instrumental polarization leakage.
\cite{moore_et_al_2013} predict a level of contamination based on simulations
of the polarized point source population at low frequencies.
Although this predicted level of contamination
may already be ruled out by measurements from \cite{bernardi_et_al_2013},
these effects are a real concern for 21~cm EoR experiments.
In the present analysis, however, we do not consider this contamination;
rather, we
assume that the dense $uv$ coverage of our concept array will allow
for precision calibration and image-based primary beam correction not possible
with the sparse PAPER array.  Through careful and concerted effort this
systematic should be able to be reduced to below the EoR level.

As discussed in \S\ref{sec:sense_calc}, we do consider the ``multi-baseline
mode mixing" effects presented in \cite{hazelton_et_al_2013}.  These
effects may result when partially coherent baselines are combined
to improve power spectrum sensitivity, introducing
additional spectral structure in the foregrounds and thus 
complicating their mitigation. Conversely, the fact that only instantaneously
redundant baselines were combined in \cite{pober_et_al_2013b} and
\cite{parsons_et_al_2013} was partially responsible for the clear
separation between the wedge and EoR window.  
Since recent, competitive upper limits were set using this 
conservative approach, we include it as our ``pessimistic" 
foreground strategy, noting that recent progress in accounting for 
the subtleties in  partially coherent analyses \citep{hazelton_et_al_2013} 
make it likely that better schemes will be available soon.

To encompass all these uncertainties in the foreground emission and foreground
removal techniques, we use three models for our foregrounds, which we refer to
in shorthand as ``pessimistic," ``moderate," and ``optimistic".
These models are summarized in Table \ref{tab:fg}.
\begin{table}
\centering
\begin{tabular}{p{.75in}|p{2in}} Model & Parameters \\
\hline
Moderate & Foreground wedge extends $0.1\ h{\rm Mpc}^{-1}$ beyond horizon limit \\
Pessimistic & Foreground wedge extends $0.1\ h{\rm Mpc}^{-1}$ beyond horizon limit, and only instantaneously redundant baselines can be combined coherently \\
Optimistic & Foreground wedge extends to FWHM of primary beam 
\end{tabular}
\caption{Summary of the three foreground removal models.}
\label{tab:fg}
\end{table}
 
The ``moderate" model is chosen to
closely mirror the predictions and data from
PAPER.  In this model the wedge is considered to extend 
$\Delta k_{\parallel} = 0.1\ h{\rm Mpc}^{-1}$ beyond the horizon limit.
The exact scale of the ``horizon+.1" limit to the wedge is motivated
by the predictions of 
\cite{parsons_et_al_2012b} and the measurements of 
\cite{pober_et_al_2013b} and
\cite{parsons_et_al_2013}.   Although the exact extent of the ``supra-horizon"
emission (i.e. the ``+.1") at the level of the EoR signal remains to be 
determined, all of these constraints point to a range of 0.05 to 0.15
$h\rm{Mpc}^{-1}$.  The uncertainty in this value does not have a large
effect on the ultimate power spectrum sensitivity of next generation 
measurements.   
For shorthand, we will sometimes 
refer this model as having a
``horizon wedge."

The ``pessimistic" model uses the same horizon wedge as the moderate
model, but assumes that only 
instantaneously redundant baselines are coherently
combined. Any non-redundant baselines which sample the same $uv$ pixel
as another baseline --- either through being similar in length
and orientation or through the effects of earth rotation --- 
are added incoherently.
In effect, this model
covers the case where the multi-baseline mode-mixing of
\cite{hazelton_et_al_2013} cannot be corrected for.
Significant efforts are underway to develop pipelines
which correct for this effect and recover the sensitivity boost of partial
coherence;
since these algorithms have yet to be demonstrated on actual observations,
however, we consider this our pessimistic scenario.  

The final ``optimistic" model, assumes the EoR window remains workable
down to $k_{\parallel}$ modes bounded by the FWHM of the primary beam,
as opposed to the horizon: $k_{\parallel,\rm{pb}} = \sin(\rm{FWHM}/2)\times k_{\parallel,\rm{hor}}$.
The specific choice of the FWHM is somewhat arbitrary; one could also
consider a wedge extending the first-null in the primary beam (although
this is ill-defined for a Gaussian beam model).  
Alternatively, one might envision a ``no wedge" model
meant to mirror the case where
foreground removal techniques work optimally, removing all foreground
contamination down to the intrinsic spectral structure of the foreground
emission.
In practice, the small 
$\sim 10^{\circ}$ size of the HERA primary beam renders these
different choices effectively indistinguishable.
Therefore, our choice of the primary beam FWHM can also 
be considered representative
of nearly all cases where foreground removal proves highly effective.
As of the writing of this paper, no foreground removal algorithms have proven
successful to these levels, although this is admittedly a
somewhat tautological statement, since no published measurements have reached
the sensitivity level of an EoR detection.
Furthermore, the sampling point-spread function (PSF) 
in $k$-space at low $k$'s is expected to make
clean, unambiguous retrieval of these modes exceedingly difficult
\citep{liu_and_tegmark_2011,parsons_et_al_2012b}, although
the small size of the HERA primary beam ameliorates this problem
by limiting the scale of this PSF.  We find this
effect to represent a small ($\lesssim 5\%$)
correction to the low-$k$ sensitivities reported in this work.
In effect, the optimistic model is included to both show the
effects of foregrounds on the recovery of the 21~cm power spectrum, and to
give an impression of what could be achievable.  For shorthand, this model will
be referred to as having a ``primary beam wedge."

Incorporating these foreground models into the sensitivity calculations
described in \S\ref{sec:instrument} is quite straightforward.  Modes deemed
``corrupted" by foregrounds according to a model are simply excluded
from the 3D $k$-space cube, and therefore contribute no sensitivity
to the resultant power spectrum measurements.

\subsection{Reionization}
\label{sec:eor}

In order to encompass the large theoretical uncertainties in the
cosmic reionization history, we use the publicly available 
\texttt{21cmFAST}\footnote{http://homepage.sns.it/mesinger/DexM\_\_\_21cmFAST.html/}
code v1.01 \citep{mesinger_and_furlanetto_2007,mesinger_et_al_2011}.
This semi-numerical code allows us to quickly generate large-scale simulations
of the ionization field (400 Mpc on a side) 
while varying key parameters to examine the possible
variations in the 21~cm signal.  Following \cite{mesinger_et_al_2012}, we choose
three key parameters to encompass the maximum variation in the signal: 
\begin{enumerate}

\item \emph{$\zeta$, the ionizing efficiency}: $\zeta$ is a conglomeration of
a number of parameters relating to the amount of ionizing photons escaping
from high-redshift galaxies: $f_{\rm esc}$, the fraction of ionizing
photons which escape into the IGM, $f_*$, the star formation efficiency,
$N_\gamma$, the number of ionizing photons produced per baryon in stars,
and $n_{\rm rec}$ the average number of recombinations per baryon.
Rather than parameterize the uncertainty in these quantities individually,
it is common to define $\zeta = f_{\rm esc}f_*N_\gamma/(1+n_{\rm rec})$
\citep{furlanetto_et_al_2004}.  We explore a range
of $\zeta = 10-50$ in this work, which is generally consistent with current
CMB and Ly$\alpha$ constraints on reionization \citep{mesinger_et_al_2012}.

\item \emph{$T_{\rm vir}$, the minimum virial temperature of halos
producing ionizing photons}: $T_{\rm vir}$ parameterizes the mass
of the halos responsible for reionization.  Typically, $T_{\rm vir}$ is
chosen to be $10^4$~K, which corresponds to a halo mass of $10^8~\rm{M}_\odot$
at $z = 10.$  This value is chosen because it represents the temperature at
which atomic cooling becomes efficient.  In this work, we explore $T_{\rm vir}$
ranging from $10^3 \mbox{--} 3 \times 10^5$~K to span the uncertainty in high-redshift
galaxy formation physics as to which halos host significant stellar
populations (see e.g. \citealt{haiman_et_al_1996},
\citealt{abel_et_al_2002} and \citealt{bromm_et_al_2002} for lower mass limits
on star-forming halos,
and e.g. \citealt{mesinger_and_dijkstra_2008} and \citealt{okamoto_et_al_2008}
for feedback effects which can suppress low mass halo star formation).

\item \emph{$R_{\rm mfp}$, the mean free path of ionizing photons 
through the intergalactic medium (IGM)}: $R_{\rm mfp}$ sets the
maximum size of HII regions that can form during reionization.  Physically,
it is set by the space density of Lyman limit systems, which act as sinks
of ionizing photons.  In this work, we explore a range of mean free paths
from 3 to 80 Mpc, spanning the uncertainties in current measurements
of the mean free path at $z\sim 6$ \citep{songaila_and_cowie_2010}.

\end{enumerate}
We note there are many other tunable parameters that could affect
the reionization history.  In particular,
the largest 21~cm signals can be produced in models where
the IGM is quite cold during
reionization (cf. \citealt{parsons_et_al_2013}).  We do not
include such a model here, and rather focus on the potential
uncertainties within ``vanilla" reionization; for an analysis of
the detectability of early epochs of X-ray heating, see 
\cite{christian_and_loeb_2013} and \cite{mesinger_et_al_2013}.
Also note that \texttt{21cmFAST} assumes the values of the EoR
parameters are constant over all redshifts considered.
With the exception of our three EoR variables, we use the fiducial parameters
of the \texttt{21cmFAST} code; see \cite{mesinger_et_al_2011} for more details.

Note we do assume that $T_{\rm spin} \gg T_{\rm CMB}$ at all epochs, which
could
potentially create a brighter signal at the highest redshifts.  Given
that thermal noise generally dominates the signal at the highest redshifts
regardless, we choose to ignore this effect, noting that it will only
increase the difficulties of $z > 10$ observations we describe below.
(Although this situation may be changed by the 
alternate X-ray heating scenarios considered in 
\citealt{mesinger_et_al_2013}.)

\subsubsection{``Vanilla" Model}

\begin{figure}[ht!]
\centering
\includegraphics[width=3.5in]{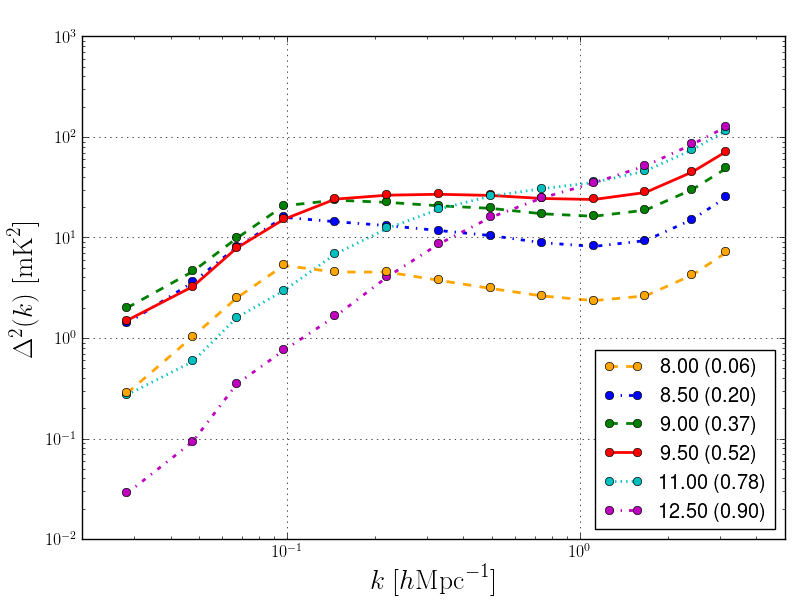}
\caption{Power spectra at several redshifts for our vanilla
reionization model with $\zeta = 31.5,\ T_{\rm vir} = 1.5 \times 10^4~\rm{K},$ 
and $R_{\rm mfp} = 30~\rm{Mpc}$.
Numbers in parentheses give the neutral fraction at that redshift.}
\label{fig:pspec_vanilla}
\end{figure}

For the sake of comparison, it is worthwhile to have one fiducial model
with ``middle-ground" values for all the parameters in question.  
We refer to this model as our ``vanilla" model.  Note that this
model was not chosen because we believe it most faithfully
represents the true reionization history of the universe
(though it is consistent with current observations).  Rather,
it is simply a useful point of comparison for all the other realizations
of the reionization history.
In this model, the values of the three parameters being studied are
$\zeta = 31.5,\ T_{\rm vir} = 1.5 \times 10^4~\rm{K}$ and $R_{\rm mfp} = 30~\rm{Mpc}$.
This model achieves 50\% ionization at $z\sim 9.5$, and complete ionization
at $z\sim7$.
The redshift evolution of the power spectrum in this model is shown in Figure
\ref{fig:pspec_vanilla}.

\subsubsection{The Effect of the Varying the EoR Parameters}
\label{sec:varying}

\begin{figure*}[ht!]
\centering
\includegraphics[width=6.5in]{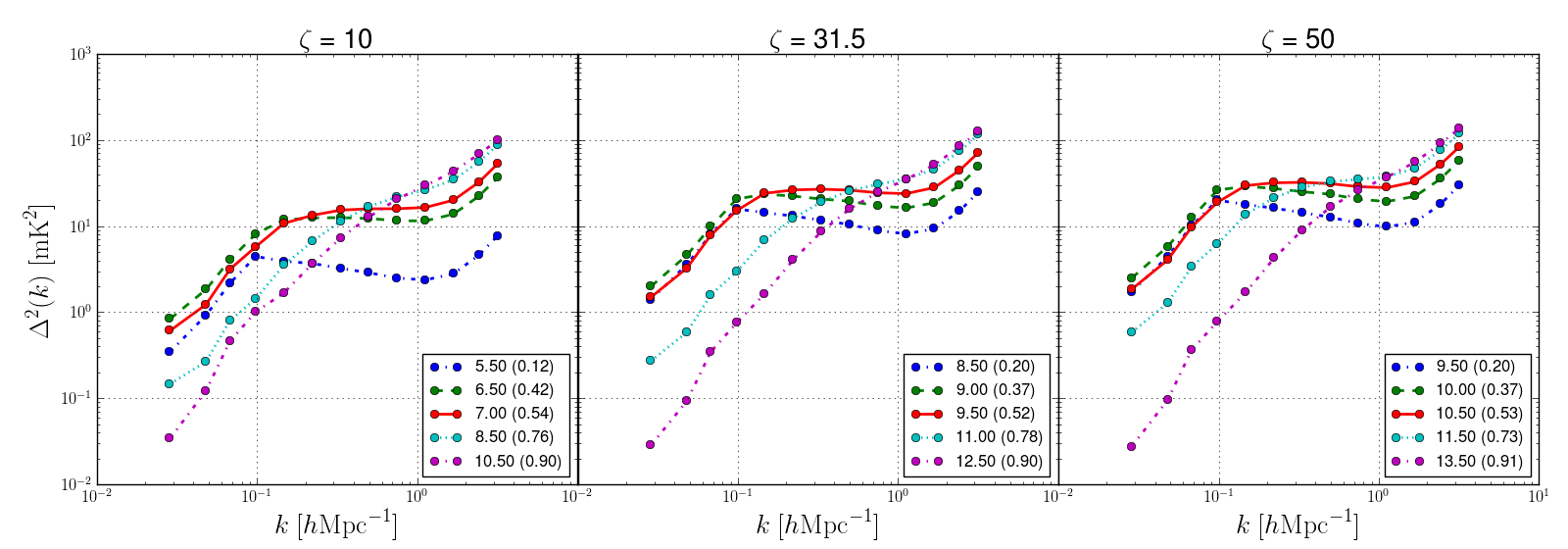}
\includegraphics[width=6.5in]{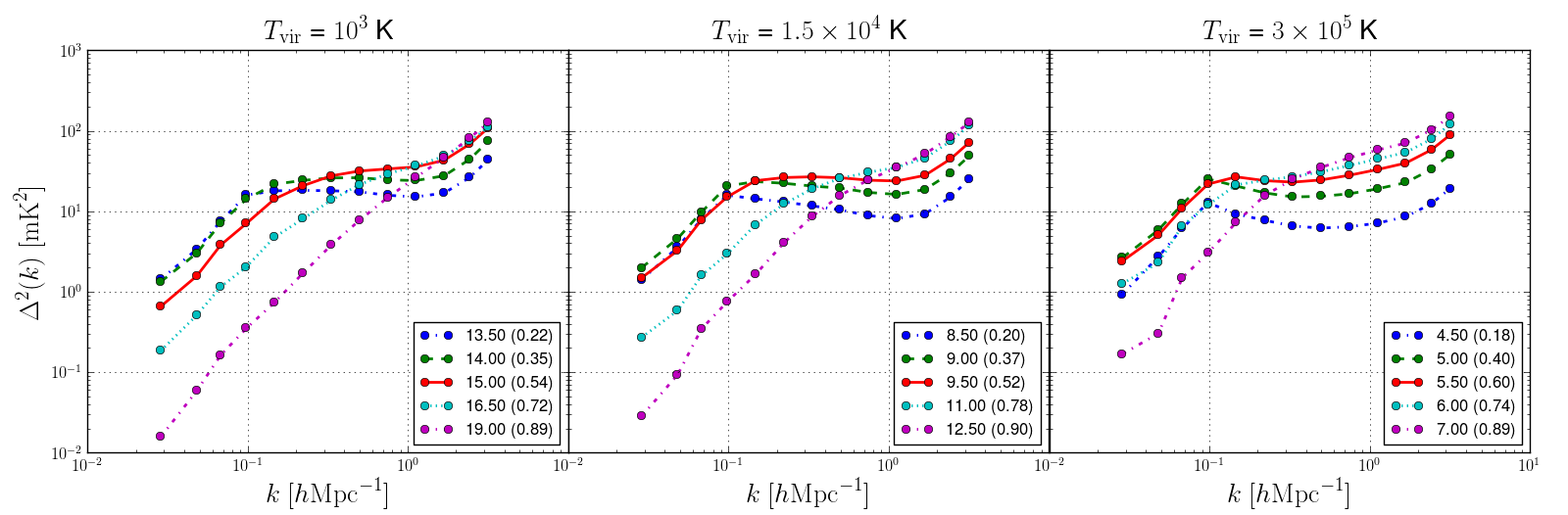}
\includegraphics[width=6.5in]{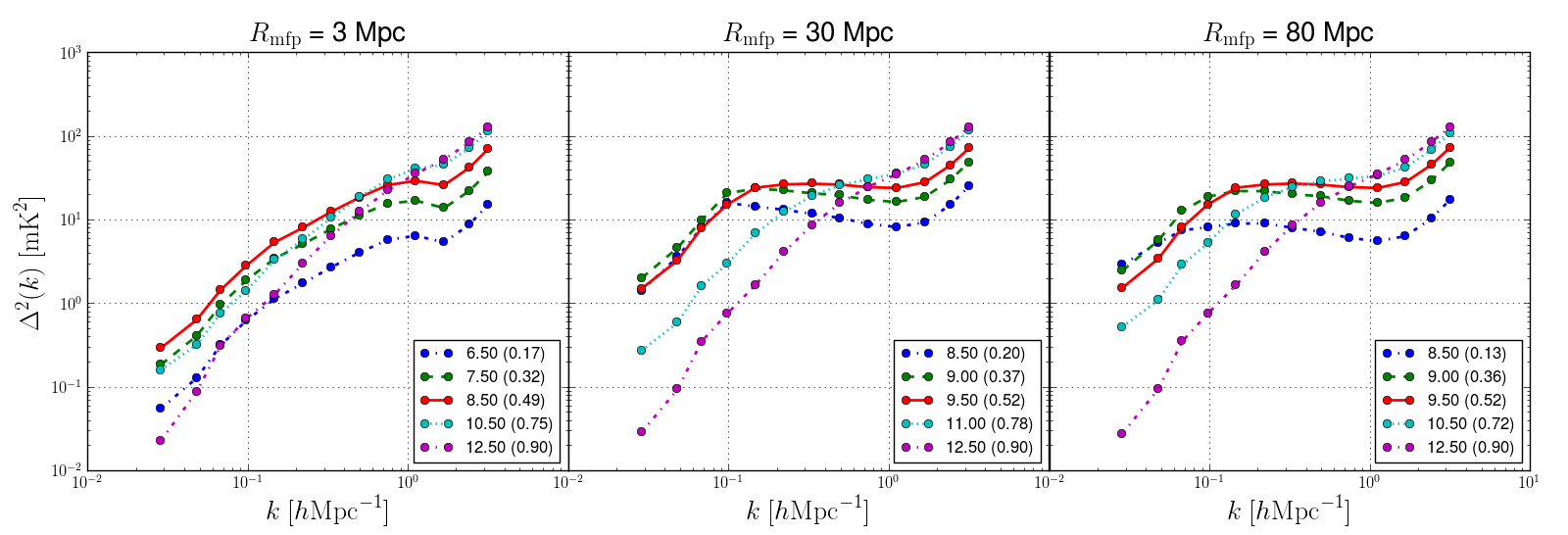}
\caption{Power spectra as a function of redshift for the low, high,
and fiducial
values of the ionizing efficiency, $\zeta$ (top row), $T_{\rm vir}$ (middle
row) and $R_{\rm mfp}$ (bottom row).  Exact values of each
parameter are given in the panel title.
Numbers in parentheses give the neutral fraction at that redshift.
The central panel is the vanilla model and is identical to Figure
\ref{fig:pspec_vanilla} (although the $z=8$ curve is not included for
clarity).  Colors in each panel map to roughly the same
neutral fraction.  Qualitative effects of varying each parameter are
apparent: $\zeta$ changes the timing of reionization but not the shape
of the power spectrum; $T_{\rm vir}$ drastically alters the timing
of reionization with smaller effects on the power spectrum shape; and
small values of $R_{\rm mfp}$ reduce the amount of low $k$ power.}
\label{fig:pspecs}
\end{figure*}

The effects of varying $\zeta$, $T_{\rm vir}$ and $R_{\rm mfp}$ are
illustrated in Figure \ref{fig:pspecs}.
Each row shows the effect of varying one of the three parameters while
holding the other two fixed.
The middle panel in each
row is for our vanilla model, and thus is the same as Figure 
\ref{fig:pspec_vanilla} (although the $z=8$ curve is not included for
clarity).
Several qualitative observations can immediately be made.  Firstly, we can see
from the top row that
$\zeta$ does not significantly change the shape of the power spectrum, but only the duration
and timing of reionization.  This is expected, since the same sources
are responsible for driving reionization regardless of $\zeta$.  Rather,
it is only the number of ionizing photons that these sources produce that
varies.

Secondly, we can see from the middle row that the most dramatic
effect of $T_{\rm vir}$ is to substantially change the timing of reionization.
Our high and low values of $T_{\rm vir}$ create reionization histories
that are inconsistent with current constraints from the CMB and Ly$\alpha$
forest \citep{fan_et_al_2006,hinshaw_et_al_2012}.  This alone does not rule out these values of $T_{\rm vir}$ for the
minimum mass of reionization galaxies, but it does mean that some additional 
parameter would have to be adjusted within our vanilla model to create
reasonable reionization histories.  We can also see 
that the halo virial temperature affects the shape of the power spectrum.  When
the most massive halos are responsible for reionization, we see significantly
more power on very large scales than in the case where low-mass galaxies
reionize the universe.

Finally, the bottom row shows that the mean free path of ionizing
photons also affects the amount of large scale power in the 21~cm power spectrum.
$R_{\rm mfp}$ values of 30 and 80~Mpc produce essentially
indistinguishable power spectra,
except at the very largest scales at late times.  However, the very small
value of $R_{\rm mfp}$ completely changes the shape of the power spectrum, 
resulting in a steep slope versus $k$, even at
50\% ionization, where most models show a fairly flat power spectrum up to
a ``knee" feature on larger scales.  In section \ref{sec:distinguishing}, we
consider using some of these characteristic features to qualitatively assess
properties from reionization in 21~cm measurements.

\section{Results}
\label{sec:results}

In this section, we will present the predicted sensitivities
that result from combinations of EoR and foreground models.  
We will focus predominantly on the moderate model
where one can take advantage of partially-redundant baselines,
but the wedge still contaminates $k_{\parallel}$
modes below $0.1~h\rm{Mpc}^{-1}$ above the horizon limit.`
In presenting the sensitivity levels 
achievable under the other two foreground models, we focus on the
additional science that will be prevented/allowed if these models represent
the state of foreground removal analysis.

We will take several fairly orthogonal approaches towards understanding
the science that will be achievable.
First, in \S\ref{sec:sensitivity},
our approach is to attempt to cover the broadest range
of possible power spectrum shapes and amplitudes in order to make
generic conclusions about the detectability of the 21~cm power spectrum.
In \S\ref{sec:timing}, \S\ref{sec:imaging}, and 
\S\ref{sec:distinguishing}, we focus on our vanilla reionization model
and semi-quantitatively explore the physical lessons the predicted sensitivities
will permit.
Finally, in \S\ref{sec:adrian}, we undertake a Fisher matrix analysis
and focus on specific determinations of EoR parameters with respect to the
fiducial evolution of our vanilla model, exploring the degeneracies between
parameters and providing lessons in how to break them.
The end result of these various analyses is a holistic picture about the
kinds of information we can derive from next generation EoR measurements.

\subsection{Sensitivity Limits}
\label{sec:sensitivity}

In this section, we consider the signal-to-noise ratio of
power spectrum measurements 
achievable under our various foreground removal models.
The main results are presented in Figures \ref{fig:pspec-errs},
\ref{fig:hor-snr-matrix}, \ref{fig:inst-red-snr-matrix},
and \ref{fig:pb-snr-matrix}.
Figure \ref{fig:pspec-errs}
shows the constraints on the 50\% ionization power spectrum
in our vanilla model
for each of the three foreground models, as well as the measurement
significances of alternate ionization histories using the vanilla model.
Figures \ref{fig:hor-snr-matrix}, \ref{fig:inst-red-snr-matrix},
and \ref{fig:pb-snr-matrix} show the power spectrum measurement
signifances that result when the EoR parameters are varied
for the moderate, pessimistic, and optimistic foreground models 
respectively.
 
\subsubsection{Methodology}
In order to explore the largest range of possible power spectrum shapes and amplitudes, it is important to keep in mind the small but non-negligible spread between various theoretical predictions in the literature.  To avoid having to run excessive numbers of simulations, we make use of the observation that much of the differences between simulations is due to discrepancies in their predictions for the ionization history $x_\textrm{HI} (z)$, in the sense that the differences decrease if neutral fraction (rather than redshift) is used as the time coordinate for cross-simulation comparisons.  We thus make the ansatz that given a single set of parameters $(\zeta,T_{\rm vir},R_{\rm mfp})$, the \texttt{21cmFAST} power spectrum outputs can (modulo an overall normalization factor) be translated in redshift to mimic the outputs of alternative models that predict a different ionization history.
In practice, the \texttt{21cmFAST} simulation provides a suite
of power spectra in either (a) fixed steps in $z$ or (b) approximately
fixed steps in $x_{\rm HI}$, but constrained to appear at a single $z$. 
We utilize the latter set,
and ``extrapolate" each neutral fraction to a variety
of redshifts by scaling the amplitude with the square of the 
mean temperature of the IGM
as $(1+z)$, as anticipated when ionization fractions
dominate the power spectrum \citep{mcquinn_et_al_2006,lidz_et_al_2008}.
While not completely motivated by the physics 
of the problem (since within \texttt{21cmFAST} a given set of EoR parameters
does produce only one reionization history), this approach allows
us to explore an extremely wide range of power spectrum amplitudes
while running a reasonable number of simulations.

\subsubsection{Moderate Foreground Model}

\begin{figure*}[ht!]
\centering
\includegraphics[width=3.25in]{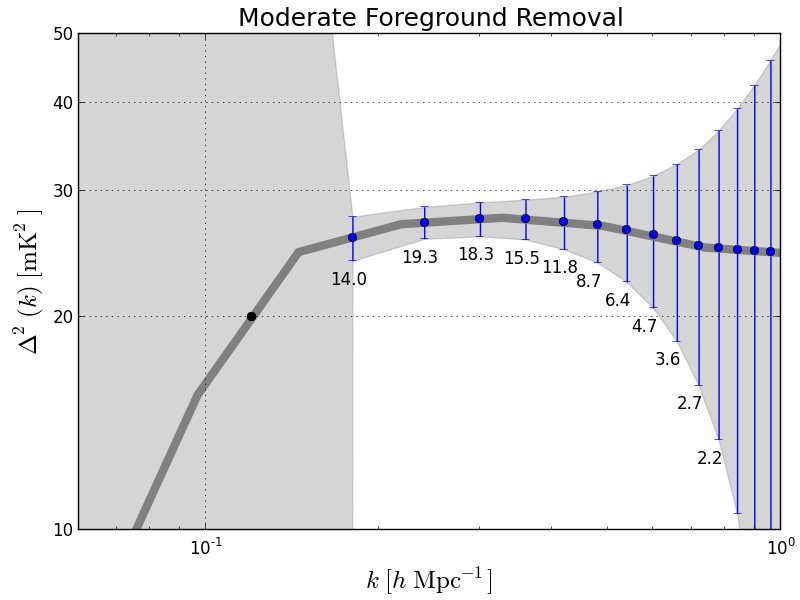}\includegraphics[width=3.5in]{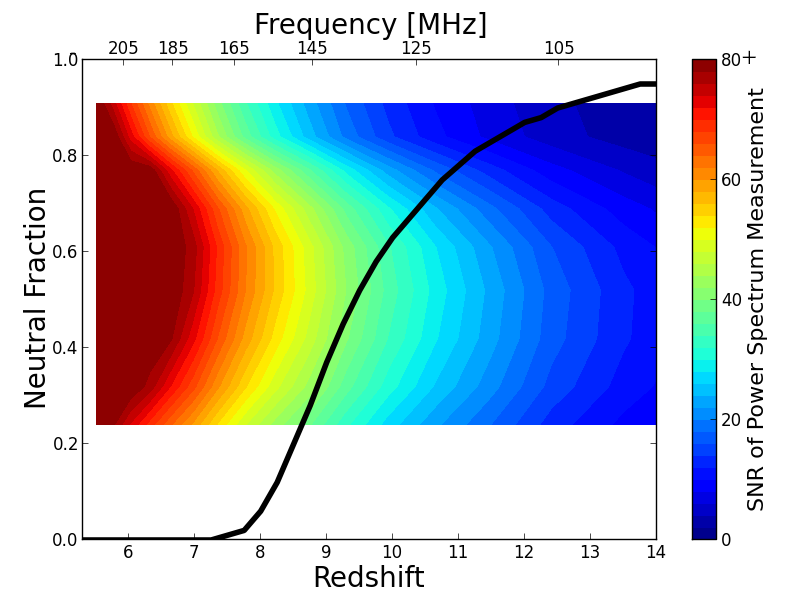}

\includegraphics[width=3.25in]{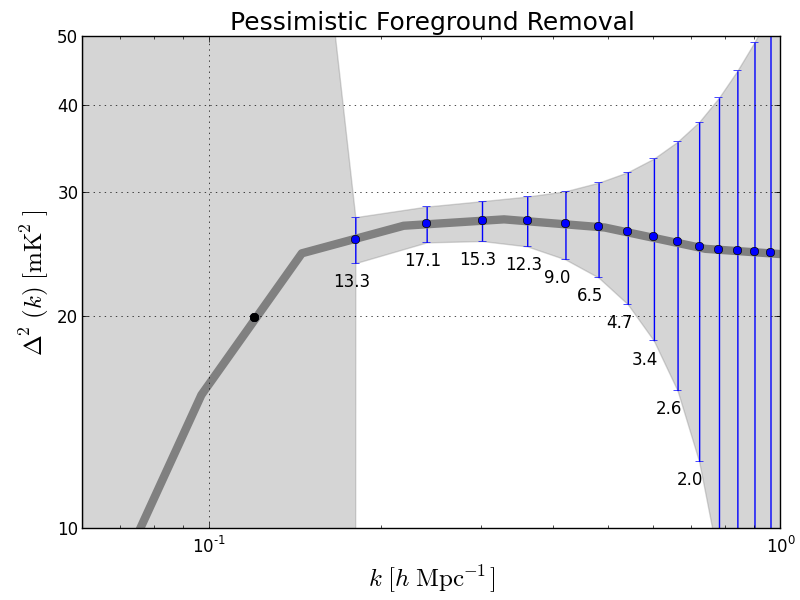}\includegraphics[width=3.5in]{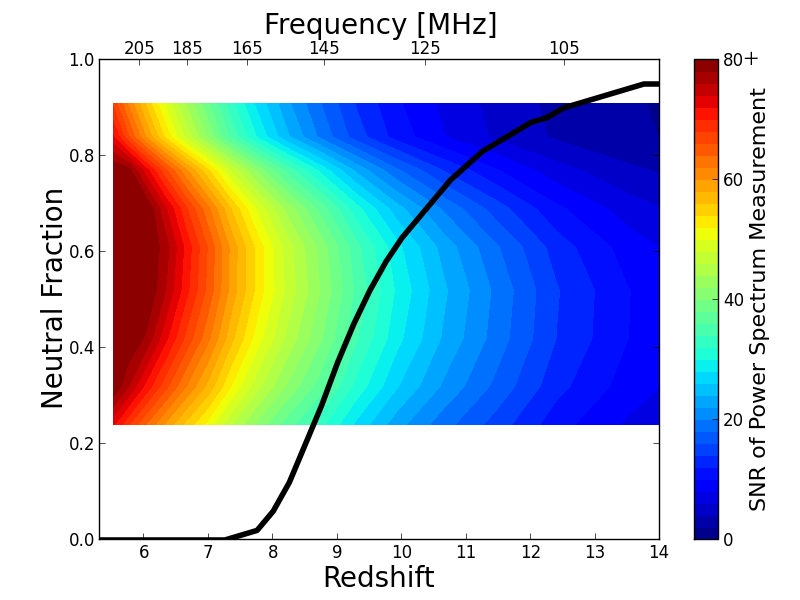}

\includegraphics[width=3.25in]{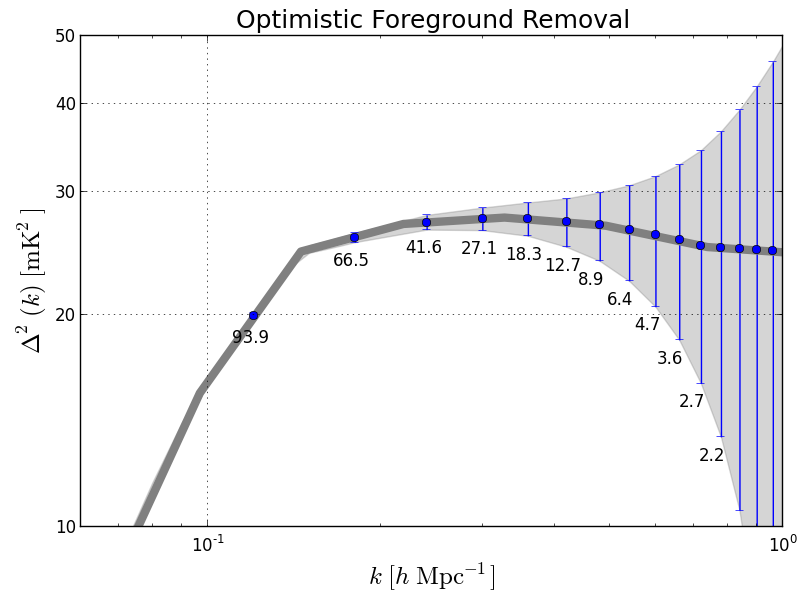}\includegraphics[width=3.5in]{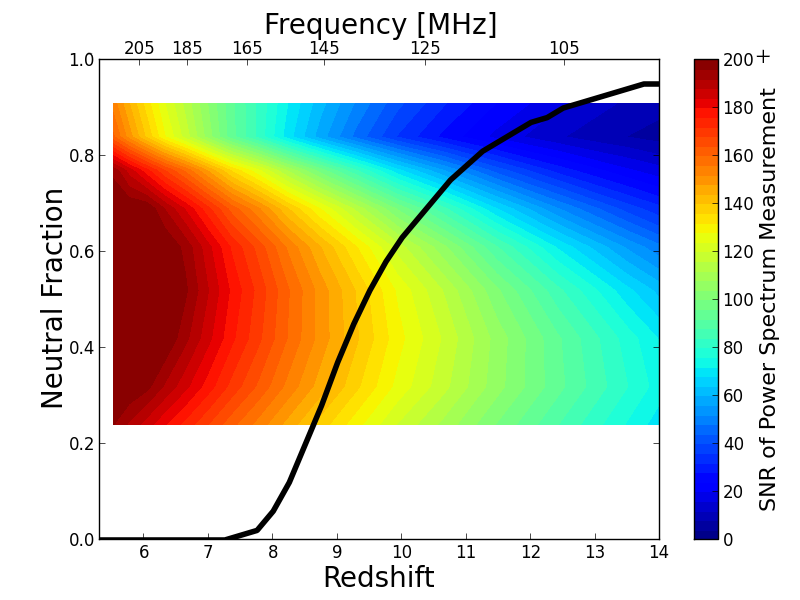}
\caption{
\emph{Left:} Power spectrum constraints on the fiducial EoR model
at $z = 9.5\ (53\%$ ionization) for each of the three foreground removal models:
moderate (top), pessimistic (middle) and optimistic (bottom).
The shaded gray region shows the $1\sigma$ error range, whereas the
location of the blue error bars indicate the binned sampling pattern; the binning
is set by the bandwidth of 8 MHz.
Black points without error bars
indicate measurements allowed by instrumental parameters, but
rendered unusable according to the foreground model.  
The net result of
these measurements are $38\sigma$, $32\sigma$, and $133\sigma$
detections of the fiducial power spectrum for the moderate, pessimistic
and optimistic models, respectively.
Individual numbers below each error bar indicate the significance of the
measurement in that bin.
\emph{Right:} Colored contours show the total SNR of a power spectrum detection 
as a function of redshift and
neutral fraction for the three foreground removal models:
moderate (top), pessimistic (middle) and optimistic (bottom).
The black curve shows the fiducial evolution of the 
vanilla model; contour values off of the black curve are obtained by translating the
fiducial model in redshift.  This figure therefore allows one to examine
the SNR for a far broader range of reionization histories than only those
predicted by simulations with vanilla model parameters.  Alternative evolution
histories are less physically motivated, since a given set of EoR parameters
does only predict one evolution history.  
The plotted sensitivities assume 8 MHz bandwidths are used to measure
the power spectra, so not all points in the horizontal direction are 
independent.  The incomplete coverage versus $x_{\rm HI}$ does not indicate
that measurements cannot be made at these neutral fractions; rather, it
is a feature of the \texttt{21cmFAST} code, and is explained in 
\S\ref{sec:sensitivity}.
}
\label{fig:pspec-errs}
\end{figure*}

Figure \ref{fig:pspec-errs} shows forecasts for constraints on our
fiducial reionization model under the three foreground scenarios.
The left-hand panels of the figure show the constraints on the
spherically averaged power
spectrum at $z = 9.5$, the point of 50\% ionization in this model, for 
the three foreground removal models.
(The 50\% ionization point generally corresponds to the peak power
spectrum brightness at the scales of interest --- as can be seen
in Figure \ref{fig:pspecs} --- making its detection
a key goal of reionization experiments \citep{lidz_et_al_2008,
bittner_and_loeb_2011}.) For the moderate model (top row), the
errors amount to a $38\sigma$ detection of the 21~cm power spectrum at 
50\% ionization. 

\begin{figure*}[ht!]
\centering
\includegraphics[width=5.75in]{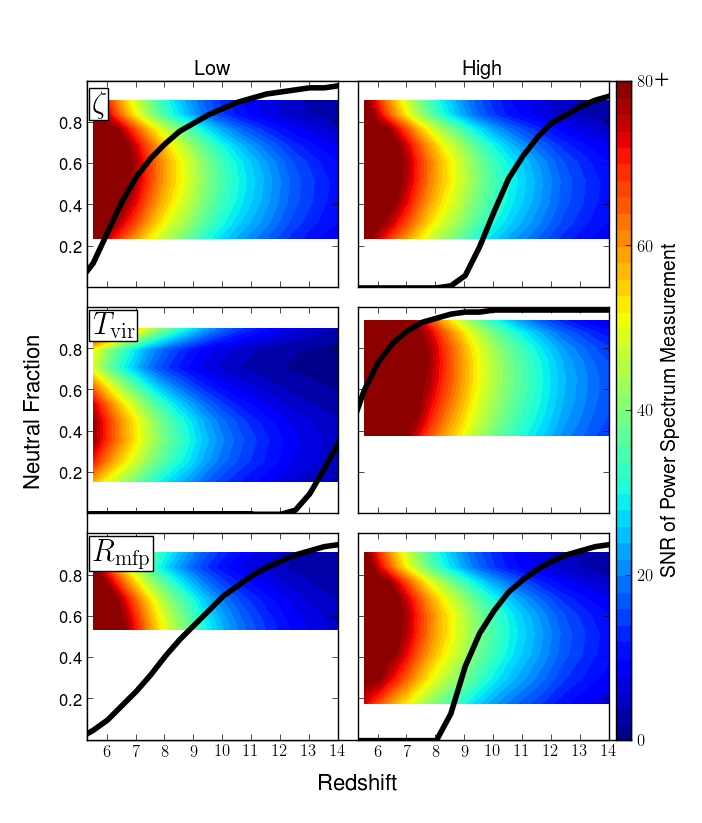}
\caption{Signal-to-noise ratio of 21~cm power spectrum detections 
under the moderate foreground scenario for the high and low
values of the parameters in our EoR models as functions of neutral
fraction and redshift.  In each panel,
one parameter is varied, while the other two are held fixed
at the ``vanilla" values.  The black curve shows the
fiducial evolution for that set of model parameters.
The incomplete coverage versus $x_{\rm HI}$ does not indicate
that measurements cannot be made at these neutral fractions; rather, it
is a feature of the \texttt{21cmFAST} code, and is explained in 
\S\ref{sec:sensitivity}.
\emph{Top}: the ionizing efficiency, $\zeta$.  
Values are $\zeta = 10$ (left) and $\zeta = 50$ (right).
\emph{Middle}: the minimum virial temperature of ionizing haloes, $T_{\rm vir}$.
Values are $T_{\rm vir} = 1\times 10^3$~K (left) and $T_{\rm vir} = 3 \times 10^5$~K (right).
\emph{Bottom}: the mean free path for ionizing photons through the IGM, $R_{\rm mfp}$.
Values are $R_{\rm mfp} = 3$~Mpc (left) and $R_{\rm mfp} = 80$~Mpc (right).
The moderate foreground removal scenario generically
allows for a high significance measurement for nearly any reasonable 
reionization history.
}
\label{fig:hor-snr-matrix}
\end{figure*}

The right-hand panels of Figure \ref{fig:pspec-errs}
warrant somewhat detailed explanation.  The three rows
again correspond to the three foreground removal models.  In each panel,
the horizontal axis
shows redshift and the vertical axis shows neutral fraction; thus this space
spans a wide range of possible reionization histories.  The black
curve is the evolution of the vanilla model through this space. 
The colored contours show
the signal-to-noise ratio of a HERA measurement of the 21~cm power spectrum
at that point in redshift/neutral fraction space, where the power
spectrum of a given $x_{\rm HI}$ is extrapolated in redshift space
as described in the beginning of \S\ref{sec:sensitivity}.
The colorscale is set to saturate at different values in each row:
$80\sigma$ (moderate and pessimistic) and $200\sigma$ (optimistic).
These sensitivities assume 8 MHz bandwidths are used to measure
the power spectra, so not every value on the redshift-axis can be taken as
an independent measurement.
The non-uniform coverage versus ionization fraction (i.e. the white
space at high and low values of $x_{\rm HI}$) --- which appears with different
values in the panels of Figures \ref{fig:pspec-errs}, \ref{fig:hor-snr-matrix}, 
\ref{fig:inst-red-snr-matrix}, and \ref{fig:pb-snr-matrix} --- 
is a feature of the \texttt{21cmFAST} code when
attempting to produce power spectra for a set of input parameters at relatively
even spaced values of ionization fraction.  
The black line is able to extend into the white region because it 
was generated to have uniform spacing in $z$ as opposed to $x_{\rm HI}$.
The fact that these values are
missing has minimal impact on the conclusions drawn in this work.

In the moderate model, the 50\% ionization point of the fiducial power spectrum
evolution is detected at $\sim40\sigma$.  However, we see that
nearly every ionization
fraction below $z \sim 9$ is detected with equally high significance.
In general, the contours follow nearly vertical lines through this space.
This implies that the evolution of sensitivity as a function of redshift
(which is primarily driven by the system temperature) is much stronger
than the evolution of power spectrum amplitude as a function of neutral fraction
(which is primarily driven by reionization physics).

Figure \ref{fig:hor-snr-matrix} shows signal-to-noise contour plots
for six different variations of our EoR parameters, using only the
moderate foreground scenario.  (The pessimistic and optimistic equivalents
of this figure are shown in Figures \ref{fig:inst-red-snr-matrix}
and \ref{fig:pb-snr-matrix}, respectively.)
In each panel,
we have varied one parameter from the fiducial vanilla model.  In particular,
we choose the lowest and highest values of each parameter considered in
\S\ref{sec:eor}.  Since we
extrapolate each power spectrum to a wide variety of redshifts, choosing
only the minimum and maximum values leads to little loss of generality.
Rather, we are picking extreme shapes and amplitudes for the power spectrum,
and asking whether they can be detected if such a power spectrum were
to correspond to a particular redshift.  And, as with the vanilla model
shown in Figure \ref{fig:pspec-errs},
it is clear that the moderate foreground removal scenario
allows for the 21~cm power spectrum
to be detected with very high significance
below $z\sim 8-10$, depending on the EoR model.
Relative to the
effects of system temperature, then, the actual brightness of the power
spectrum as a function of neutral fraction plays a small role in determining
the detectability of the cosmic signal.  Of course, there is still a wide
variety of power spectrum brightnesses; for a given
EoR model, however, the relative power spectrum amplitude evolution as a 
function of redshift is fairly small.

There are also several more specific points about Figure 
\ref{fig:hor-snr-matrix} that warrant comment.  Firstly, as stated in
\S\ref{sec:varying}, the ionizing efficiency $\zeta$ has
little effect on the shape of the power spectrum, but only on the timing and
duration of reionization.  This is clear from the identical sensitivity
levels for both values of $\zeta$, as well as for the vanilla model shown
in Figure \ref{fig:pspec-errs}.  Secondly, we reiterate that by tuning
values of $T_{\rm vir}$ alone, we can produce ionization histories that
are inconsistent with observations of the CMB
and Lyman-$\alpha$ forest.  In our analysis here, we extrapolate
the power spectrum shapes produced by these extreme histories to more
reasonable redshifts to show the widest range of possible scenarios.
The fact that the fiducial evolution histories (black lines)
of the $T_{\rm vir}$ row are wholly unreasonable is understood, and does
not constitute an argument against this type of analysis.  

\subsubsection{Other Foreground Models}

It is clear then, that the moderate foreground removal scenario
will permit high sensitivity measurements of the 21~cm power with
the next generation of experiments.  Before considering what types of science
these sensitivities will enable, it is worthwhile to consider the effects
of the other foreground removal scenarios.

\begin{figure*}[ht!]
\centering
\includegraphics[width=5.75in]{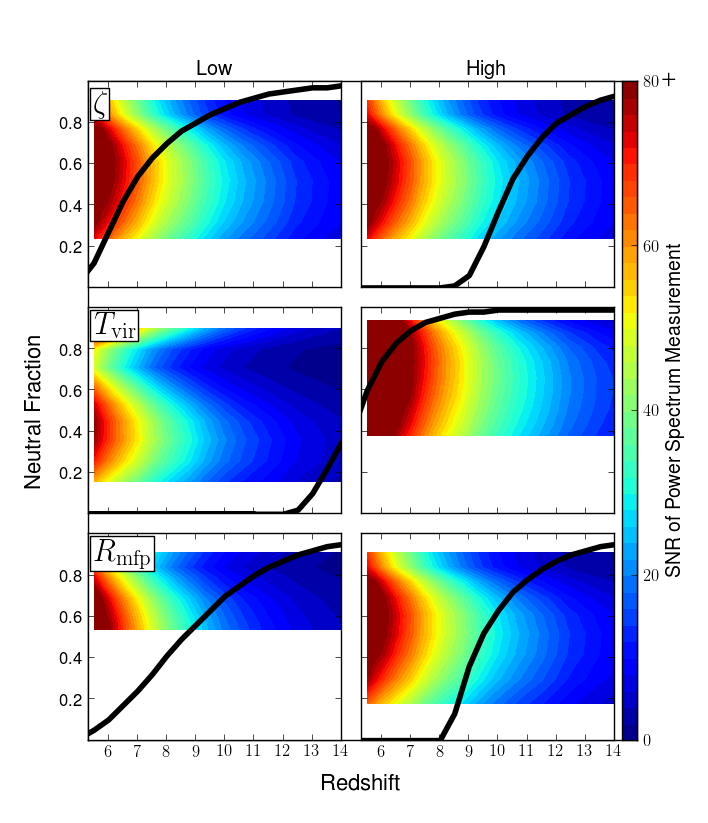}
\caption{
Same as Figure \ref{fig:hor-snr-matrix}, but for the pessimistic
foreground model.
Note that the color-scale is the same as Figure
\ref{fig:hor-snr-matrix} and saturates at $80\sigma$.}
\label{fig:inst-red-snr-matrix}
\end{figure*}

Our pessimistic scenario assumes
--- like the moderate scenario --- that
foregrounds irreparably corrupt $k_{\parallel}$ modes within the horizon
limit plus $0.1~h{\rm Mpc}^{-1}$, but also conservatively omits the 
coherent addition of partially redundant baselines in an effort to avoid
multi-baseline systematics.
As stated, this is the most
conservative foreground case we consider.
The achievable constraints on
our fiducial vanilla power spectrum under this model
were shown in the second row of Figure \ref{fig:pspec-errs}; Figure
\ref{fig:inst-red-snr-matrix} shows the sensitivities for other EoR models.
\label{fig:pspec-inst-red}
The sensitivity loss associated with coherently adding only instantaneously
redundant baselines is fairly small, $\sim 20\%$.  It should be noted
that this pessimistic model affects only the thermal noise error bars
relative to the moderate model; sample variance contributes the same
amount of uncertainty in each bin.  In an extreme sample variance limited
case, the pessimistic and moderate models would yield the same power
spectrum sensitivities.  We will further explore 
the contribution of sample variance
to these measurements in \S\ref{sec:imaging}.  Here we note that the
pessimistic model generally increases the \emph{thermal noise} uncertainties by
$30\mbox{--}40\%$ over the moderate model.  This effect will be greater
for an array with less instantaneous redundancy than the HERA concept design.

\begin{figure*}[ht!]
\centering
\includegraphics[width=5.75in]{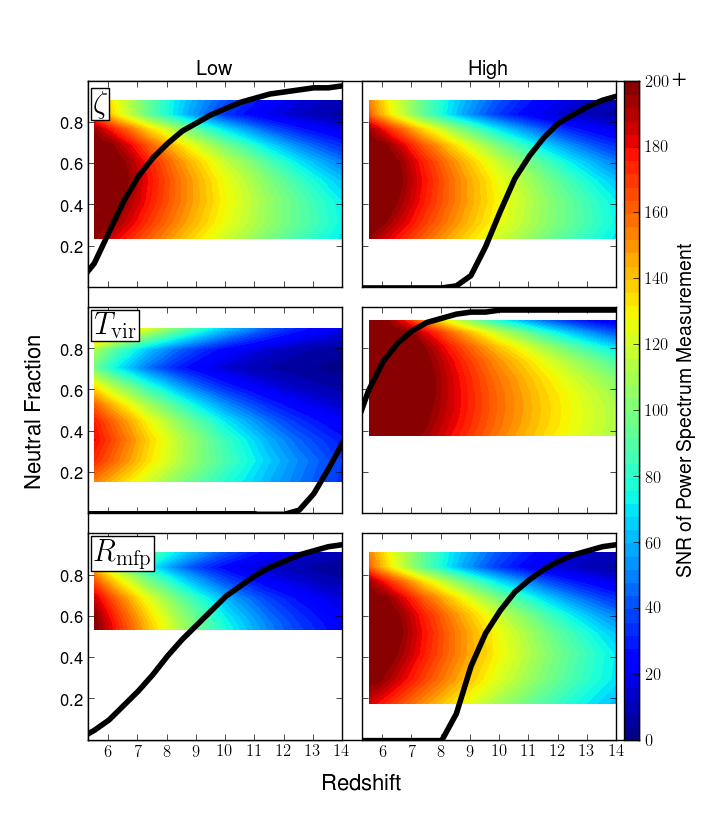}
\caption{
Same as Figures \ref{fig:hor-snr-matrix} and \ref{fig:inst-red-snr-matrix}, 
but for the optimistic
foreground model.
Note that the color-scale has changed to saturate at $200\sigma$.}
\label{fig:pb-snr-matrix}
\end{figure*}

Finally, the sensitivity to the vanilla EoR model under the optimistic
foreground removal scenario is shown
in the bottom row of Figure \ref{fig:pspec-errs}.
Figure \ref{fig:pb-snr-matrix} shows the
sensitivity results for the other EoR scenarios.
The sensitivities for the optimistic model are exceedingly high.  
Comparison of the top and bottom rows of
Figure \ref{fig:pspec-errs} shows that this model does not uniformly
increase sensitivity across $k$-space, but rather the gains are entirely
at low $k$s.  This behavior is expected, since the effect of the optimistic
model is to recover large scale modes that are treated as foreground contaminated
in the other models.  The sensitivity boosts come from the fact that thermal 
noise is very low at these large scales, since noise scales as $k^3$ 
while the cosmological signal remains relatively flat in
$\Delta^2(k)$ space.  We consider the effect of sample variance in these
modes in \S\ref{sec:imaging}.

\subsection{The Timing and Duration of Reionization}
\label{sec:timing}

\begin{figure*}[ht!]
\centering
\includegraphics[width=7in]{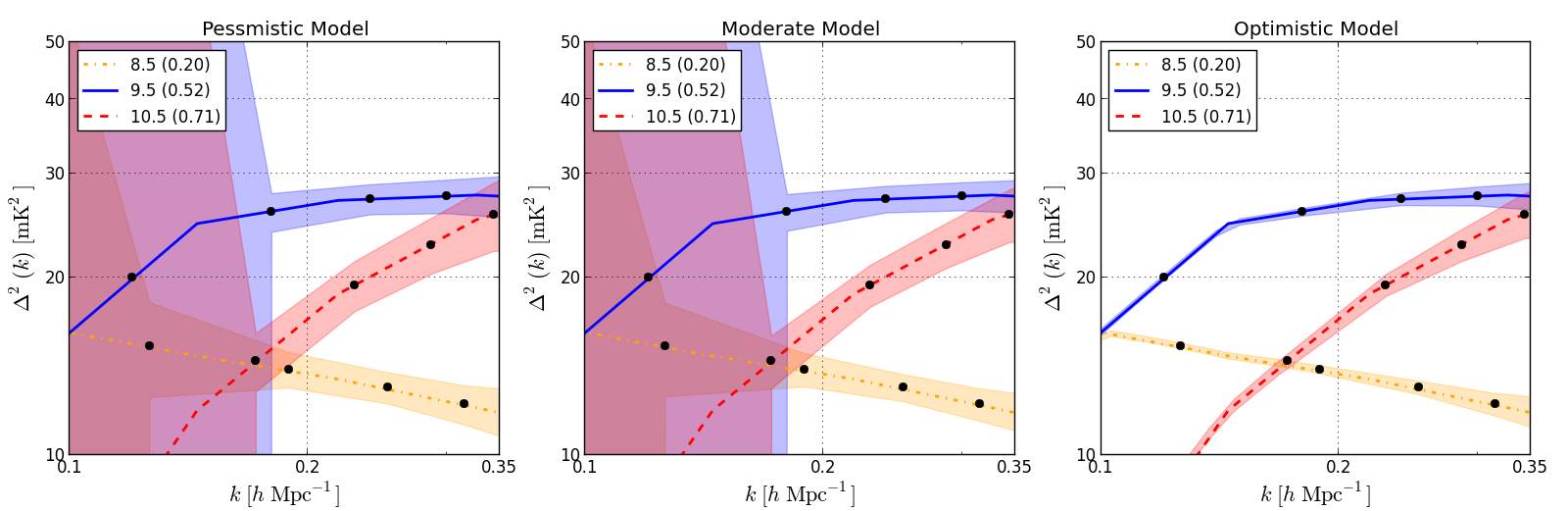}
\caption{
$1\sigma$ uncertainties in the measurements of the fiducial EoR power 
spectrum at redshifts 8.5, 9.5 and
10.5, corresponding to neutral fractions of 0.20, 0.52 and 0.71, 
respectively.
Different panels show the results for the different foreground models:
pessimistic (\emph{left}), moderate (\emph{middle}), and optimistic 
(\emph{right}).  The pessimistic and moderate scenarios should both permit 
measurements of $\Delta z\sim1.0$.  The optimistic scenario
will allow for detailed characterization of the power spectrum evolution.
}
\label{fig:zpeak}
\end{figure*}

One of the first key parameters that is expected from 21~cm measurement of the
EoR power spectrum is the redshift at which the universe was 50\% ionized,
sometimes referred to as ``peak reionization."  The rationale behind this expectation
is evident from Figure \ref{fig:pspecs}, 
where the power spectrum generically achieves peak 
brightness at $k\sim0.1~h{\rm Mpc}^{-1}$ for $x_{\rm HI} = 0.5$.  However,
given the steep increase of $T_{\rm sys}$, one must ask if an experiment
will truly have the sensitivity to distinguish the power spectrum 
at $z_{\rm peak}$ from those on either side.  
Figure \ref{fig:zpeak}
shows the error bars on our fiducial power spectrum model at 50\% ionization
($z = 9.5$), as well as those on the neighboring redshifts $z = 8.5$ and
$z = 10.5$, under each of our three foreground models.
In both the pessimistic and moderate models (left and middle panel),
the $z=8.5$ (20\% neutral), $z=9.5$ (52\% neutral)
and $z=10.5$ (71\% neutral) are distinguishable at the few sigma level.
This analysis therefore suggests that it should
be possible to identify peak reionization to within a $\Delta z \sim 1$,
with a strong dependence on the actual redshift of reionization (since
noise is significantly lower at lower redshifts).  

It is worth noting, however, that even relatively high significance
detections of the power spectrum ($\gtrsim 5\mbox{--}10\sigma$) 
may not permit one to distinguish power spectrum of peak reionization from
those at nearby redshifts --- especially as one looks to higher $z$.
For our vanilla EoR model, we find a $\sim10\sigma$ detection is necessary
to distinguish the $z = 8.5,\ 9.5,$ and 10.5 power spectra at the $>1\sigma$
level.  In fact, for this level of significance, nearly all of the power spectra at redshifts higher
than peak reionization at $z = 9.5$ are indistinguishable given the steep
rise in thermal noise.  Even if the current generation of 21~cm telescopes does yield
a detection of the 21~cm power spectrum, these first measurements
do not guarantee stringent constraints on the peak
redshift of reionization.

Finally, one can see that the high sensitivities permitted 
by the optimistic foreground model
will allow a detailed characterization of the power spectrum amplitude and slope
as a function of redshift.  We discuss exactly what kind
of science this will enable
(beyond detecting the timing and duration of reionization)
in \S\ref{sec:adrian}.

\subsection{Sample Variance and Imaging}
\label{sec:imaging}

\begin{figure}[ht!]
\centering
\includegraphics[width=3.3in]{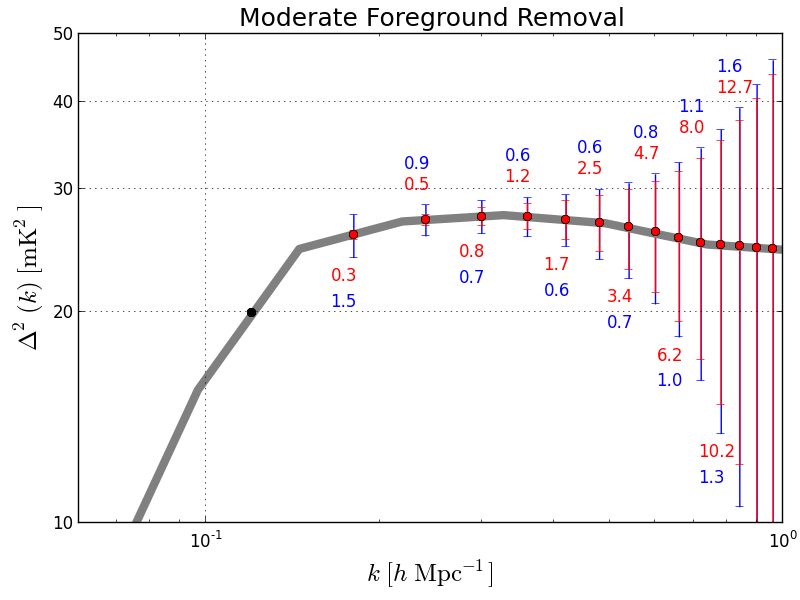}
\caption{The breakdown of the error bars in the top-left panel of Figure \ref{fig:pspec-errs} (vanilla EoR model and moderate foreground removal scenario).
Red shows the contribution of thermal noise, blue the contribution of sample
variance.  The text shows the value of each contribution in mK$^2$ --- 
\emph{not} the significance of the detection, as in previous plots.
Regions where the sample variance error dominates the thermal noise error
are in the imaging regime.  The placement of the numerical values above
or below the error bar has no significance; it is only for clarity.
Sample variance dominates the errors in the moderate foreground scenario
on scales $k < 0.25~h{\rm Mpc}^{-1}$.}
\label{fig:samp-var-hor}
\end{figure}

Given the high power spectrum sensitivities achievable under all of our
foreground removal models, one
must investigate the contributions of sample variance to the overall errors.
For the moderate foreground model, 
Figure \ref{fig:samp-var-hor} shows the relative contribution of sample variance
and thermal noise to the errors shown in 
the top-left panel of Figure \ref{fig:pspec-errs}.
From this plot, it is clear that sample variance contributes over half of
the total power spectrum uncertainty on scales $k < 0.3 h{\rm Mpc}^{-1}$.  If
the power spectrum constituted the ultimate measurement of reionization, this
would be an argument for a survey covering a larger area of sky.  For our
HERA concept array, which drift-scans, this is not possible, but may be
for phased-array designs.  However, the sample variance dominated regime
is very nearly equivalent to the imaging regime: thermal noise is reduced
to the point where individual modes have an SNR $\gtrsim$ 1.  Therefore, using a
filter to remove the wedge region (e.g. \citealt{pober_et_al_2013b}),
a collecting area of $0.1~{\rm km}^2$ should provide sufficient sensitivity
to image the Epoch of Reionization over
$\sim800$ sq. deg. (6~hours of right ascension $\times\ 
8.7^{\circ}~\rm{FHWM})$ on scales of $0.1 \mbox{--} 0.25~h{\rm Mpc}^{-1}$.
We note that the HERA concept design is not necessarily optimized for imaging;
other configurations may be better suited if imaging the EoR is the primary
science goal.

The effect of the other foreground models on imaging is relatively
small.  The
poorer sensitivities of the pessimistic model push up thermal noise,
lowering the highest $k$ that can be imaged to $k \sim 0.2 h{\rm Mpc}^{-1}$.
The optimistic
foreground model recovers significant SNR on the largest scales, to the
point where sample variance dominates all scales up to 
$0.3~h{\rm Mpc}^{-1}$.  The effects of foregrounds and the wedge on imaging
with a HERA-like array will be explored in future work. 

\subsection{Characteristic Features of EoR Power Spectrum}
\label{sec:distinguishing}

Past literature has discussed two simple features of the 21~cm power
spectra to help distinguish between models: the slope of the power
spectrum and the sharp drop in power (the ``knee") on the largest
scales \citep{mcquinn_et_al_2006}. 
In particular, the mass of the halos driving reionization (parametrized
in this analysis by the minimum virial temperature of ionizing halos)
should affect the slope of the power spectrum.  Since more massive
halos are more highly clustered, they should concentrate power on
larger scales, yielding a flatter slope.
The second row of Figure \ref{fig:pspecs} suggests this effect is
small, although not implausible.
The knee of the power spectrum at large scales should correspond to the
largest ionized bubble size, since there will be little power on scales
larger than these bubbles \citep{furlanetto_et_al_2006a}.  The position
of the knee should be highly sensitive to the mean free path for ionizing
photons through the IGM, since this sets how large bubbles can grow.
This argument is indeed confirmed by the third row of Figure \ref{fig:pspecs}, where
the smaller values of $R_{\rm mfp}$ lack significant power on large
scales compared to those models with larger values.
Unfortunately, since our third parameter $\zeta$ does not change the shape
of the power spectrum, constraining different values of $\zeta$ will not be
possible with only a shape-based analysis.
In this section we first extend these qualitative arguments based on salient
features in the power spectra, and then present a more quantitative
analysis on distinguishing models in \S\ref{sec:adrian}.

To quantify the slope of the power spectrum, we fit a straight line
to the predicted power spectrum values between $k = 0.1 \mbox{--} 1.0~h{\rm Mpc}^{-1}$.
When we refer to measuring the slope, we refer to measuring the slope
of this line, given the error bars in the $k$-bins between 
$0.1 \mbox{--} 1.0~h{\rm Mpc}^{-1}$.
This choice of fit is not designed to encompass the full range
of information contained in measurements of the power spectrum shape.
Rather, the goal of this section is to find simple features of the
power spectrum that can potentially teach us about the EoR without
resorting to more sophisticated modeling.

\begin{figure*}[ht!]
\centering
\includegraphics[width=7in]{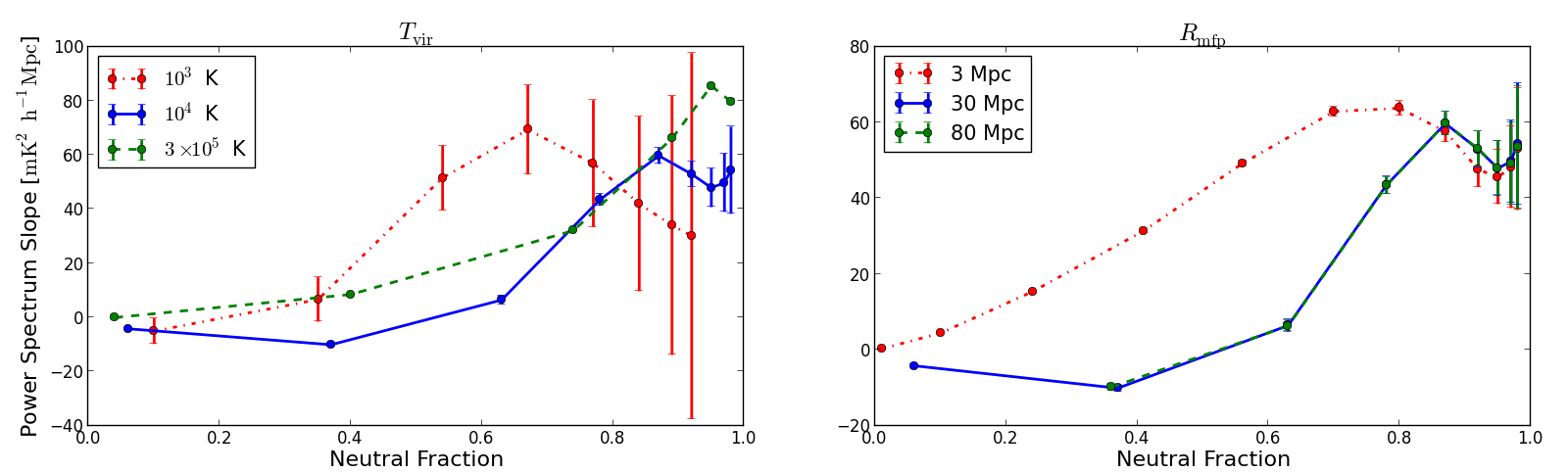}
\caption{The power spectrum slope in units
of ${\rm mK}^2 h^{-1}\rm{Mpc}$ between $k = 0.1$ and $1.0
h{\rm Mpc}^{-1}$ as a function of neutral fraction for various EoR models.
Note that error bars are plotted on all points and correspond to the redshift
of a given neutral fraction for that model.  \emph{Left}: Different values
of $T_{\rm vir}$.  \emph{Right}: Different values of $R_{\rm mfp}$.
While different values of $T_{\rm vir}$ and $R_{\rm mfp}$ produce considerable
changes in the power spectrum slope, it will be difficult to unambiguously
interpret its physical significance.}
\label{fig:slope-comp}
\end{figure*}

Figure \ref{fig:slope-comp} shows the evolution of the slope of the
linear fit to the power spectrum
over the range $k = 0.1 \mbox{--} 1.0~h{\rm Mpc}^{-1}$, as
a function of neutral fraction for several EoR models.  
Error bars in both panels correspond
to the error measured under the moderate foreground model
for a given neutral fraction in the fiducial history
of a model.  This means that, e.g., the high neutral fractions in the
$T_{\rm vir} = 10^3$ K curve have thermal noise errors corresponding to
$z \sim 20$, outside the range of many proposed experiments.  Given that
caveat, it does appear that the low-mass ionizing galaxies produce
power spectra with significantly steeper slopes at moderate neutral fractions
than those models where
only high mass galaxies produce ionizing photons.
However, it is also clear that small-bubble (i.e. low mean free path) models
can yield steep slopes.  Therefore, while there may be some physical insight
to be gleaned from measuring the slope of the power spectrum and its evolution,
even the higher sensitivity measurements permitted by the optimistic
foreground model may not be enough to break these degeneracies.
In \S\ref{sec:adrian}, we specifically focus on the kind of information
necessary to disentangle these effects.

A comparison of the error bars in the moderate and optimistic foreground
scenario measurements of the vanilla power spectrum
(rows one and three in Figure \ref{fig:pspec-errs})
reveals the difficulty in recovering the position of the knee without
foreground subtraction: foreground contamination predominantly
affects low $k$ modes, rendering large scale features like the knee
inaccessible.
In particular, the
additive component of the horizon wedge severely restricts the large scale
information available to the array.  Without probing large scales, confirming
the presence (or absence) of a knee feature is likely to be impossible.
However, Figure \ref{fig:pspec-errs} does show that if foreground removal
allows for the recovery of these modes, the knee can be detected with
very high significance, even the presence of sample variance.

\subsection{Quantitative Constraints on Model Parameters}
\label{sec:adrian}

In previous sections, we considered rather large changes to the input parameters of the \texttt{21cmFAST} model.  These gave rise to theoretical power spectra that exhibited large qualitative variations, and encouragingly, we saw that such variations should be easily detectable using next-generation instruments such as HERA.  We now turn to what would be a natural next step in data analysis following a broad-brush, qualitative discrimination: a determination of best-fit values for astrophysical parameters.  In this section, we forecast the accuracy with which $T_\textrm{vir}$, $R_\textrm{mfp}$, and $\zeta$ can be measured by a HERA-like instrument, paying special attention to degeneracies.  In many of the plots, we will omit the pessimistic foreground scenario, for often the results from it are identical to (and visually indistinguishable from) those from moderate foregrounds.  Ultimately, our results will suggest parameter constraints that are smaller than one can justifiably expect given the reasonable, but non-negligible uncertainty surrounding simulations of reionization \citep{zahn_et_al_2011}.  Our final error bar predictions (which can be found in Table \ref{finalErrors}) should therefore be interpreted cautiously, but we do expect the qualitative trends in our analysis to continue to hold as theoretical models improve.

\begin{figure*}
\centering
\includegraphics[width=1.0\textwidth,trim=2.5cm 3cm 3cm 4cm,clip]{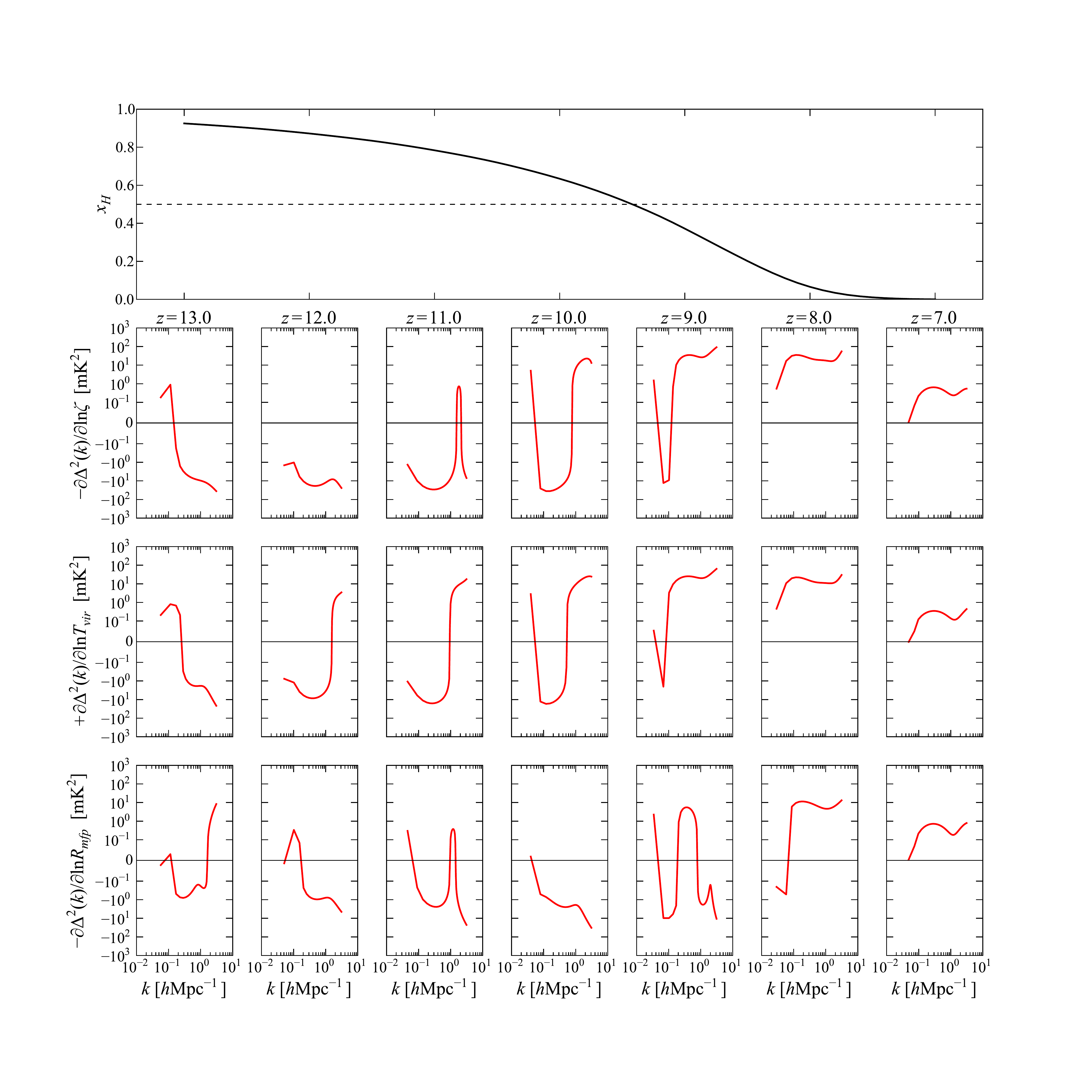}
\caption{Power spectrum derivatives as a function of wavenumber $k$ and redshift $z$.  Each of the lower three rows shows derivatives with respect to a different parameter in our three-parameter model, and the top panel (aligned in redshift with the bottom panels) shows the corresponding neutral fraction.  Because our parameter vector $\boldsymbol \theta$ (Equation [\ref{eq:FisherDef}]) contains non-dimensionalized parameters, the derivatives $\partial \Delta^2(k,z) / \partial \theta_i$ are equivalent (if evaluated at the fiducial parameter values) to the logarithmic derivatives shown here.  Note that the derivatives with respect to $\zeta$ and $R_\textrm{mfp}$ have been multiplied by $-1$ to facililate later comparisons.  The vertical axes for the derivatives are linear between $-10^{-1}$ and $10^1$, and are logarithmic outside that range.  From this figure, we see that while the lowest redshifts are easy to access observationally, the model parameters are highly degenerate.  The higher redshifts are less degenerate, but thermal noise and foregrounds make observations difficult.}
\label{fig:unweightedPk}
\end{figure*}

\subsection{Fisher matrix formalism for errors on model parameters}

To make our forecasts, we use the Fisher information matrix $\mathbf{F}$, which takes the form
\begin{equation}
\label{eq:FisherDef}
\mathbf{F}_{ij} \equiv - \Big{\langle} \frac{\partial^2 \ln \mathcal{L}}{\partial \theta_i \partial \theta_j} \Big{\rangle} = \sum_{k,z} \frac{1}{\varepsilon^2 (k,z)} \frac{\partial \Delta^2 (k,z)}{\partial \theta_i} \frac{\partial \Delta^2 (k,z)}{\partial \theta_j},
\end{equation}
where $\mathcal{L}$ is the likelihood function (i.e. the probability distribution for the measured data as a function of model parameters), $\varepsilon (k,z)$ is the error on $\Delta^2(k,z)$ measurements as a function of wavenumber $k$ and redshift $z$, and $\boldsymbol{\theta} = ( T_\textrm{vir} / T_\textrm{vir}^\textrm{fid}, R_\textrm{mfp} / R_\textrm{mfp}^\textrm{fid}, \zeta / \zeta^\textrm{fid} )$ is a vector of the parameters that we wish to measure, divided by their fiducial values\footnote{Scaling out the fiducial values of course represents no loss of generality, and is done purely for numerical convenience.}.  The second equality in Equation \eqref{eq:FisherDef} follows from assuming Gaussian errors, and  picking $k$-space and redshift bins in a way that ensures that different bins are statistically independent \citep{tegmark_et_al_1997}, as we have done throughout this paper.  Implicit in our notation is the understanding that all expectation values and partial derivatives are evaluated at fiducial parameter values.  Having computed the Fisher matrix, one can obtain the error bars on the $i^{th}$ parameter by computing $1/\sqrt{ \mathbf{F}_{ii}}$ (when all other parameters are known already) or $(\mathbf{F}^{-1})_{ii}$ (when all parameters are jointly estimated from the data).  The Fisher matrix thus serves as a useful guide to the error properties of a measurement, albeit one that has been performed optimally.  Moreover, because Fisher information is additive (as demonstrated explicitly in Equation [\ref{eq:FisherDef}]), one can conveniently examine which wavenumbers and redshifts contribute the most to the parameter constraints, and we will do so later in the section.

From Equation \eqref{eq:FisherDef}, we see that it is the derivatives of the power spectrum with respect to the parameters that provide the crucial link between measurement and theory.  If a large change in the power spectrum results from a small change in a parameter --- if the amplitude of a power spectrum derivative is large --- a measurement of the power spectrum would clearly place strong constraints on the parameter in question.  This is a property that is manifest in $\mathbf{F}$.  Also important are the shapes of the power spectrum derivatives in $(k,z)$ space.  If two power spectrum derivatives have similar shapes, changes in one parameter can be mostly compensated for by a change in the second parameter, leading to a large degeneracy between the two parameters.  Mathematically, the power spectrum derivatives can be geometrically interpreted as vectors in $(k,z)$ space, and each element of the Fisher matrix is a weighted dot product between a pair of such vectors \citep{tegmark_et_al_1997}.  Explicitly, $\mathbf{F}_{ij} = \mathbf{w}_i \cdot \mathbf{w}_j$, where
\begin{equation}
\label{eq:vecDef}
\mathbf{w}_i (k,z) \equiv \frac{1}{\varepsilon (k,z) } \frac{\partial \Delta^2 (k,z)}{\partial \theta_i}, 
\end{equation}
with the different elements of the vector corresponding to different values of $k$ and $z$.  If two $\mathbf{w}$ vectors have a large dot product (i.e. similar shapes), the Fisher matrix will be near-singular, and the joint parameter constraints given by $\mathbf{F}^{-1}$ will be poor.

\subsubsection{Single-redshift constraints}

We begin by examining how well each reionization parameter can be constrained by observations at several redshifts spanning our fiducial reionization model. In Figure \ref{fig:unweightedPk}, we show some example power spectrum derivatives\footnote{Because \texttt{21cmFAST} produces output at $k$-values that differ from those naturally arising from our sensitivity calculations, it was necessary to interpolate the outputs when computing the derivatives (which were approximated using finite-difference methods).  For this paper, we chose to fit the \texttt{21cmFAST} power spectra to sixth-order polynomials in $\ln k$, finding such a scheme to be a good balance between capturing all the essential features of the power spectrum derivatives while not over-fitting any ``noise" in the theoretical simulations.  Alternate approaches such as performing cubic splines, or fitting to fifth- or seventh-order polynomials were tested, and do not change our results in any meaningful ways.  Finally, to safeguard against generating numerical derivatives that are dominated by the intrinsic numerical errors of the simulations, we took care to choose finite-difference step sizes that were not overly fine.} as a function of $k$ and $z$.  Note that the last two rows of the figure show the \emph{negative} derivatives.  For reference, the top panel of the figure shows the corresponding evolution of the neutral fraction.  At the lowest redshifts, the power spectrum derivatives essentially have the same shape as the power spectrum itself.  To understand why this is so, note that at late times, a small perturbation in parameter values mostly shifts the time at which reionization reaches completion.  As reionization nears completion, the power spectrum decreases proportionally in amplitude at all $k$ due to a reduction in the overall neutral fraction, so a parameter shift simply causes an overall amplitude shift.  The power spectrum derivatives are therefore roughly proportional to the power spectrum.  In contrast, at high redshifts the derivatives have more complicated shapes, since changes in the parameters affect the detailed properties of the ionization field.

Importantly, we emphasize that for parameter estimation, the ``sweet spot" in redshift can be somewhat different from that for a mere detection.  As mentioned in earlier sections, the half-neutral point of $x_H = 0.5$ is often considered the most promising for a detection, since most theoretical calculations yield peak power spectrum brightness there.  This ``detection sweet spot" may shift slightly towards lower redshifts because thermal noise and foregrounds decrease with increasing frequency, but is still expected to be close to the half-neutral point.  For parameter estimation, however, the most informative redshifts may be significantly lower.  Consider, for instance, the signal at $z=8$, where $x_H = 0.066$ in our fiducial model.  Figure \ref{fig:pspec_vanilla} reveals that the power spectrum here is an order of magnitude dimmer than at $z = 9.5$ or $z = 9$, where $x_H = 0.5$ and $x_H = 0.37$ respectively.  However, from Figure \ref{fig:unweightedPk} we see that the power spectrum derivatives at $z =8$ are comparable in amplitude to those at higher redshifts/neutral fraction.  Intuitively, at $z=8$ the dim power spectrum is compensated for by its rapid evolution (due to the sharp fall in power towards the end of reionization).  Small perturbations in model parameters and the resultant changes in the timing of the end of reionization therefore cause large changes in the theoretical power spectrum.  There is thus a large information content in a $z=8$ power spectrum measurement.

\begin{figure*}[ht!]
\centering
\includegraphics[width=1.0\textwidth,trim=3cm 3cm 3cm 4cm,clip]{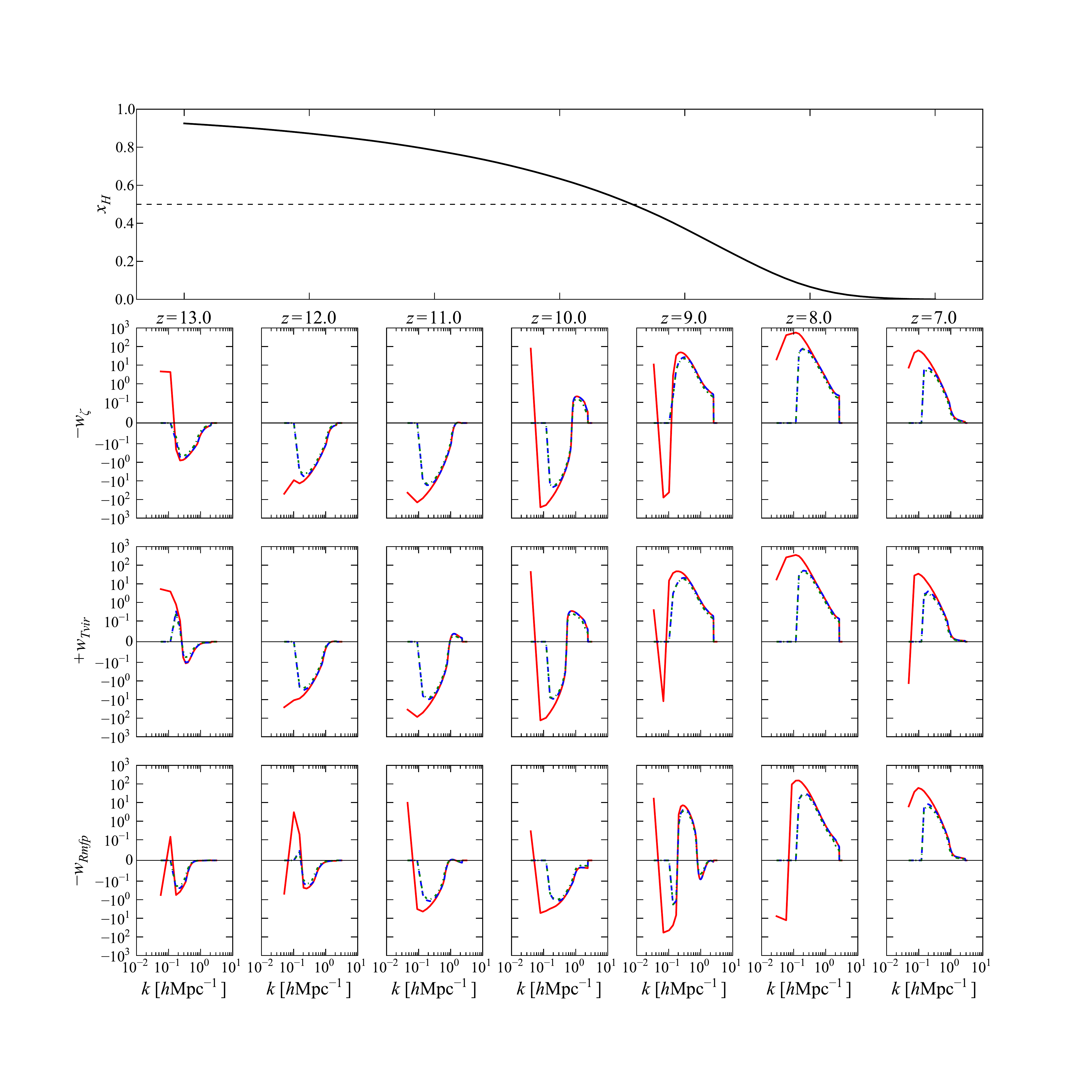}
\caption{Similar to Figure \ref{fig:unweightedPk}, but inversely weighted by the error on the power spectrum measurement, i.e. plots of $\mathbf{w}_i$ from Equation \eqref{eq:vecDef}.  These weighted derivatives are computed for HERA for the optimistic (solid red curves), moderate (dashed blue curves), and pessimistic (dot-dashed green curves) foreground models.  The pessimistic curves are essentially indistinguishable from the moderate curves.  The top panel shows the corresponding evolution of the neutral fraction.  Just as with Figure \ref{fig:unweightedPk}, the vertical axes are linear from $-10^{-1}$ to $10^1$ and logarithmic elsewhere, and $\mathbf{w}_\textrm{Rmfp}$ and $\mathbf{w}_\zeta$ have been multiplied by $-1$ to facilitate comparison with $\mathbf{w}_\textrm{Tvir}$.  With foregrounds and thermal noise, power spectrum measurements become difficult at low and high $k$ values, and constraints on model parameters become more degenerate.}
\label{fig:weightedPk}
\end{figure*}

\begin{figure*}[ht!]
\centering
\includegraphics[width=1.0\textwidth,trim=3cm 0cm 4cm 0cm,clip]{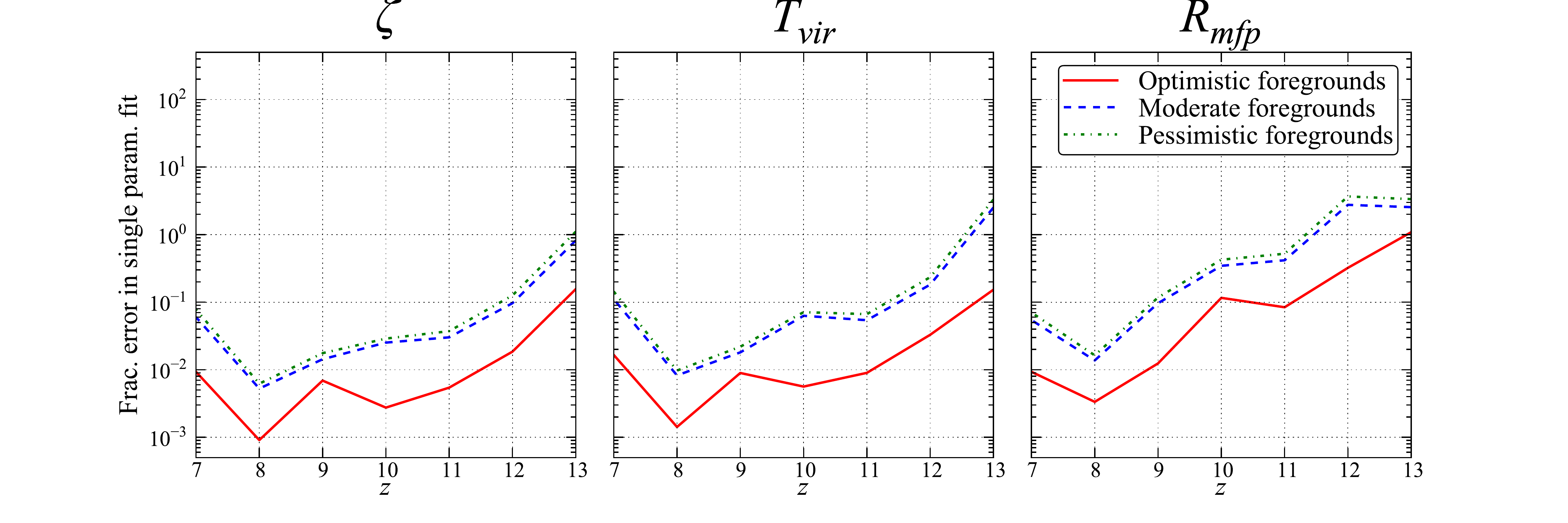}
\caption{Fractional $1\sigma$ errors ($1\sigma$ errors divided by fiducial values) as a function of redshift, with measurements at each redshift fit independently.  The errors on each parameter assume (unrealistically) that all other parameters are already known in the fit.  Solid red curves give optimistic foreground model predictions; dashed blue curves give moderate foreground model predictions; dot-dashed green curves give pessimistic foreground model predictions.  In all models, and for all three parameters, the best errors are obtained at $z=8$.  At $z=7$, the power spectrum has too small of an amplitude to yield good signal-to-noise, and at higher redshifts thermal noise poses a serious problem.}
\label{fig:singleParamErrors}
\end{figure*}

When thermal noise and foregrounds are taken into account, a $z=8$ measurement becomes even more valuable for parameter constraints than those at higher redshifts/neutral fractions.  This can be seen in Figure \ref{fig:weightedPk}, where we weight the power spectrum derivatives by the inverse measurement error\footnote{Whereas in previous sections the power spectrum sensitivities were always computed assuming a bandwidth of 8~MHz, in this section we vary the bandwidth with redshift so that a measurement centered at redshift $z$ uses all information from $z \pm 0.5$.} for HERA, producing the quantity $\mathbf{w}_i$ as defined in Equation \eqref{eq:vecDef}.  In solid red are the weighted derivatives for the optimistic foreground model, while the dashed blue curves are for the moderate foreground model.  The pessimistic case is shown using dot-dashed green curves, but these curves are barely visible because they are essentially indistinguishable from those for the moderate foregrounds.  In all cases, the derivatives peak --- and therefore contribute the most information --- at $z=8$.  Squaring and summing these curves over $k$, one can compute the diagonal elements of the Fisher matrix on a per-redshift basis.  Taking the reciprocal square root of these elements gives the error bars on each parameter assuming (unrealistically) that all other parameters are known.  The results are shown in Figure \ref{fig:singleParamErrors}.  For all three parameters, these single-parameter, per-redshift fits give the best errors at $z=8$.  At $z=7$ the neutral fraction is simply too low for there to be any appreciable signal (with even the rapid evolution unable to sufficiently compensate), and at higher redshifts, thermal noise and foregrounds become more of an influence.

Of course, what is not captured by Figure \ref{fig:singleParamErrors} is the reality that one must fit for all parameters simultaneously (since none of our three parameters are currently strongly constrained by other observational probes).  In general, our ability to constrain model parameters is determined not just by the amplitudes of the power spectrum derivatives and our instrument's sensitivity, but also by parameter degeneracies.  As an example, notice that at $z=7$, all the power spectrum derivatives shown in Figure \ref{fig:unweightedPk} have essentially identical shapes up to a sign.  This means that shifts in one parameter can be (almost) perfectly nullified by a shift in a different parameter; the parameters are degenerate.  These degeneracies are inherent to the theoretical model, since they are clearly visible even in Figure \ref{fig:unweightedPk}, where the power spectrum derivatives are shown without the instrumental weighting.  With this in mind, we see that even though Figure \ref{fig:singleParamErrors} predicts that observing the power spectrum at $z=7$ alone would give reasonable errors if there were somehow no degeneracies (making a single parameter fit the same as a simultaneous fit), such a measurement would be unlikely to yield any useful parameter constraints in practice.  To only a slightly lesser extent, the same is true for $z=8$, where the degeneracy between $R_\textrm{mfp}$ and the other parameters is broken slightly, but $T_\textrm{vir}$ and $\zeta$ remain almost perfectly degenerate.

\begin{figure*}[ht!]
\centering
\includegraphics[width=1.0\textwidth,trim=0cm 0cm 0cm 0cm,clip]{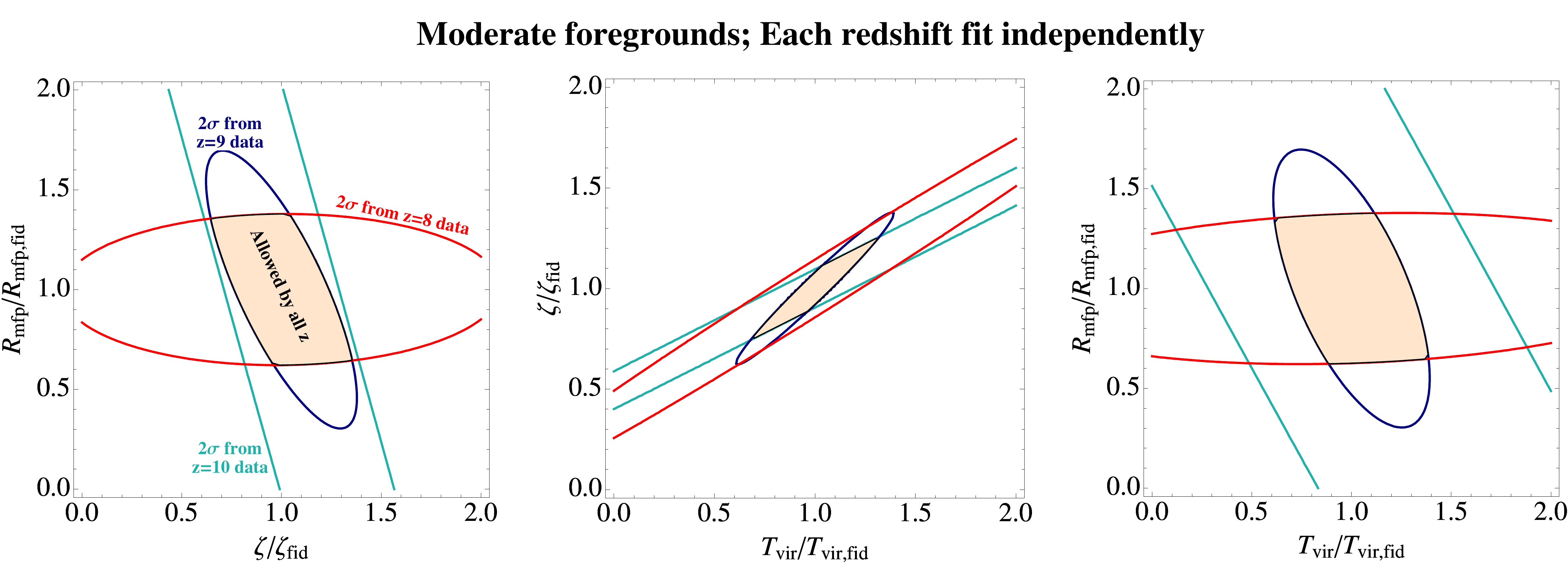}\\
\vspace{0.1in}
\includegraphics[width=1.0\textwidth,trim=0cm 0cm 0cm 0cm,clip]{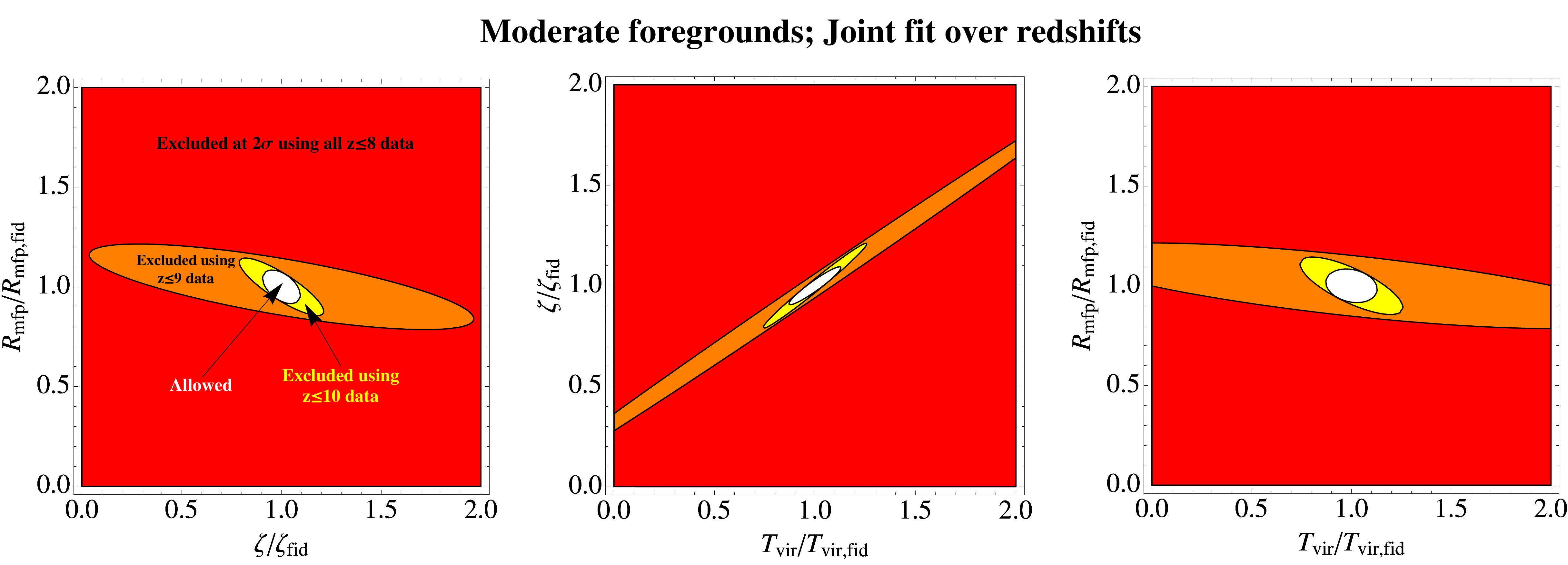}
\caption{Pairwise parameter constraints for the moderate foreground model, shown as $2\sigma$ exclusion regions.  For each pair of parameters, the third parameter has been marginalized over.  The top row shows $2\sigma$ constraints from each redshift when fit independently; the light orange regions are not ruled out by data from any redshifts.  The bottom row shows the constraints from a joint fit over multiple redshifts.  Each color represents a portion of parameter space that can be \emph{excluded} by including data up to a certain redshift.  The white ``allowed" region represents the final constraints from including all measured redshifts.  In both cases, a $z=7$ measurement alone does not provide any non-trivial constraints, but helps with degeneracy-breaking in the joint redshift fits (bottom row).  As one moves to higher and higher redshifts, power spectrum measurements probe different astrophysical processes, resulting in a shift in the principal directions of the exclusion regions.  Including higher redshifts tightens parameter constraints, but no longer helps beyond $z=10$ due to increasing thermal noise.}
\label{fig:MIDellipses}
\end{figure*}

The situation becomes even worse when one realizes that measurements at low and high $k$ are difficult due to foregrounds and thermal noise, respectively.  Many of the distinguishing features between the curves in Figure \ref{fig:unweightedPk} were located at the extremes of the $k$ axis, and from Figure \ref{fig:weightedPk}, we see that such features are obliterated by an instrumental sensitivity weighting (particularly for the pessimistic/moderate foregrounds).  This increases the level of degeneracy.  As an aside, note that with the lowest and highest $k$ values cut out, the bulk of the information originates from $k \sim 0.05~h\textrm{Mpc}^{-1}$ to $\sim 1~h\textrm{Mpc}^{-1}$ for the optimistic model and $k \sim 0.2~h\textrm{Mpc}^{-1}$ to $\sim 1~h\textrm{Mpc}^{-1}$ for the pessimistic and moderate models. (Recall again that since elements of the Fisher matrix are obtained by taking pairwise products of the rows of Figure \ref{fig:weightedPk} and summing over $k$ and $z$, the square of each weighted power spectrum derivative curve provides a rough estimate for where information comes from.)  Matching these ranges to the fiducial power spectra in Figure \ref{fig:pspec_vanilla} confirms the qualitative discussion presented in Section \ref{sec:distinguishing}, where we saw that slope of the power spectrum from $k \sim 0.1~h\textrm{Mpc}^{-1}$ to $\sim 1~h\textrm{Mpc}^{-1}$ is potentially a useful source of information regardless of foreground scenario, but that the ``knee" feature at $k \lesssim 0.1~h\textrm{Mpc}^{-1}$ will likely only be accessible with optimistic foregrounds.  This is somewhat unfortunate, for a comparison of Figures \ref{fig:unweightedPk} and \ref{fig:weightedPk} reveals that measurements at low and high $k$ would potentially be powerful breakers of degeneracy, were they observationally feasible.

\subsubsection{Breaking degeneracies with multi-redshift observations}

Absent a situation where the lowest and highest $k$ values can be probed, the only way to break the serious degeneracies in the high signal-to-noise measurements at $z=7$ and $z=8$ is to include higher redshifts, even though thermal noise and foreground limitations dictate that such measurements will be less sensitive.  Higher redshift measurements break degeneracies in two ways.  First, one can see that at higher redshifts, the power spectrum derivatives have shapes that are both more complicated and less similar, thanks to non-trivial astrophysics during early to mid-reionization.  Second, a joint fit over multiple redshifts can alleviate degeneracies even if the parameters are perfectly degenerate with each other at every redshift when fit on a per-redshift basis.  Consider, for example, the weighted power spectrum derivatives for the moderate foreground model in Figure \ref{fig:weightedPk}.  For both $z=7$ and $z=8$, the derivatives for all three parameters are identical in shape; at both redshifts, any shift in the best-fit value of a parameter can be compensated for by an appropriate shift in the other parameters without compromising the goodness-of-fit.  At $z=8$, for instance, a given fractional increase in $\zeta$ can be compensated for by a slightly larger decrease in $R_\textrm{mfp}$, since $w_\textrm{Rmfp}$ has a slightly larger amplitude than $w_\zeta$.  However, this only works if the redshifts are treated independently.  If the data from $z=7$ and $z=8$ are jointly fit, the aforementioned parameter shifts would result in a worse overall fit, because $w_\textrm{Rmfp}$ and $w_\zeta$ have roughly equal amplitudes at $z=7$, demanding fractionally equal shifts.  In other words, we see that because the \emph{ratios} of different weighted parameter derivatives are redshift-dependent quantities, joint-redshift fits can break degeneracies even when the parameters would be degenerate if different redshifts were treated independently.  It is therefore crucial to make observations at a wide variety of redshifts, and not just at the lowest ones, where the measurements are easiest.

\begin{figure*}[ht!]
\centering
\includegraphics[width=1.0\textwidth,trim=0cm 0cm 0cm 0cm,clip]{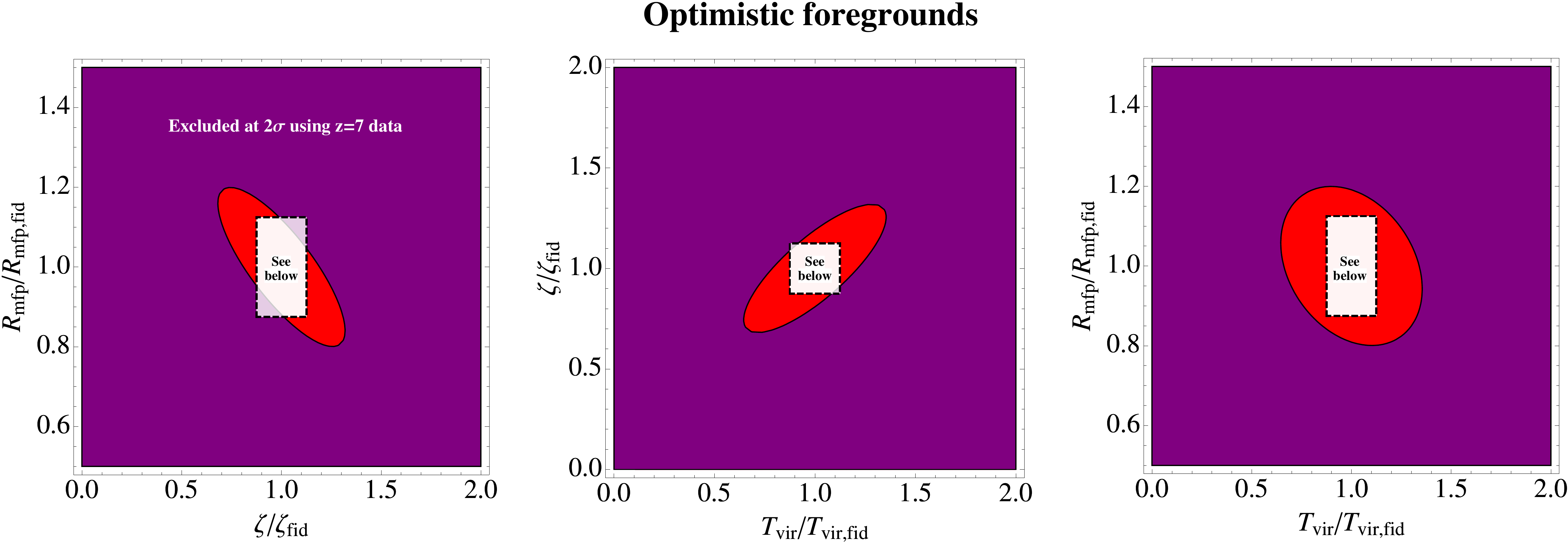}\\
\vspace{0.1in}
\includegraphics[width=1.0\textwidth,trim=0cm 0cm 0cm 0cm,clip]{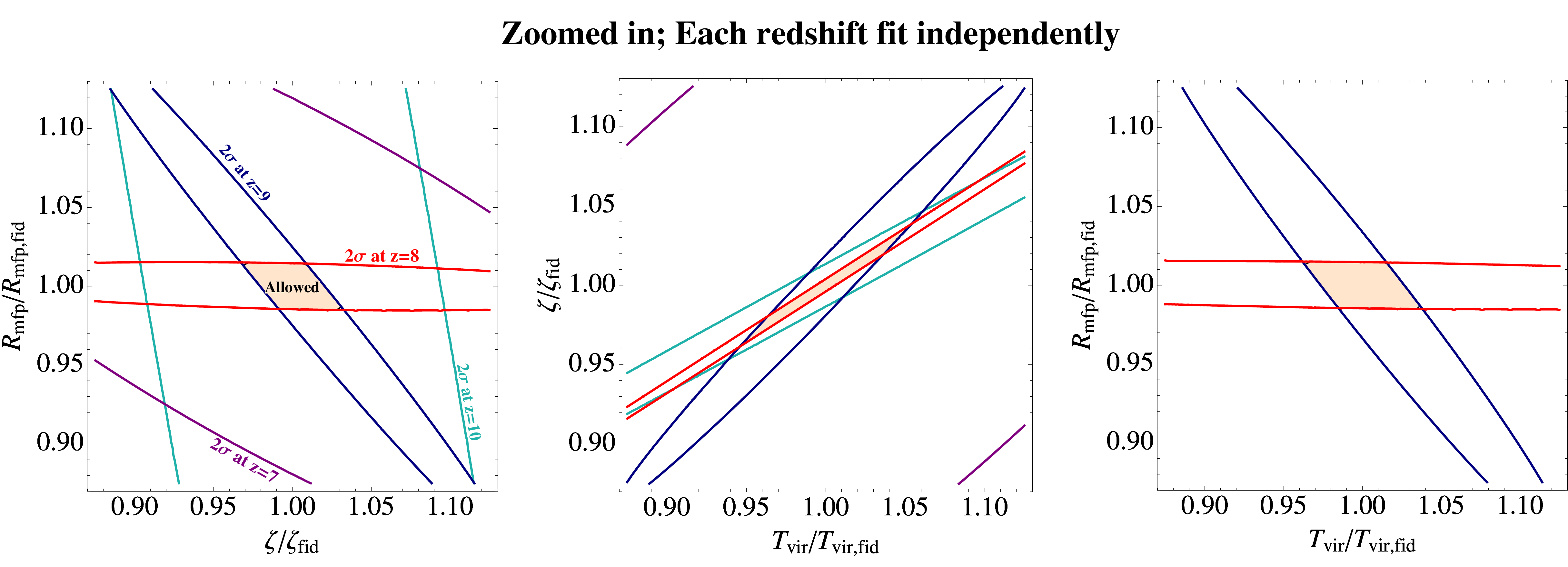}\\
\vspace{0.1in}
\includegraphics[width=1.0\textwidth,trim=0cm 0cm 0cm 0cm,clip]{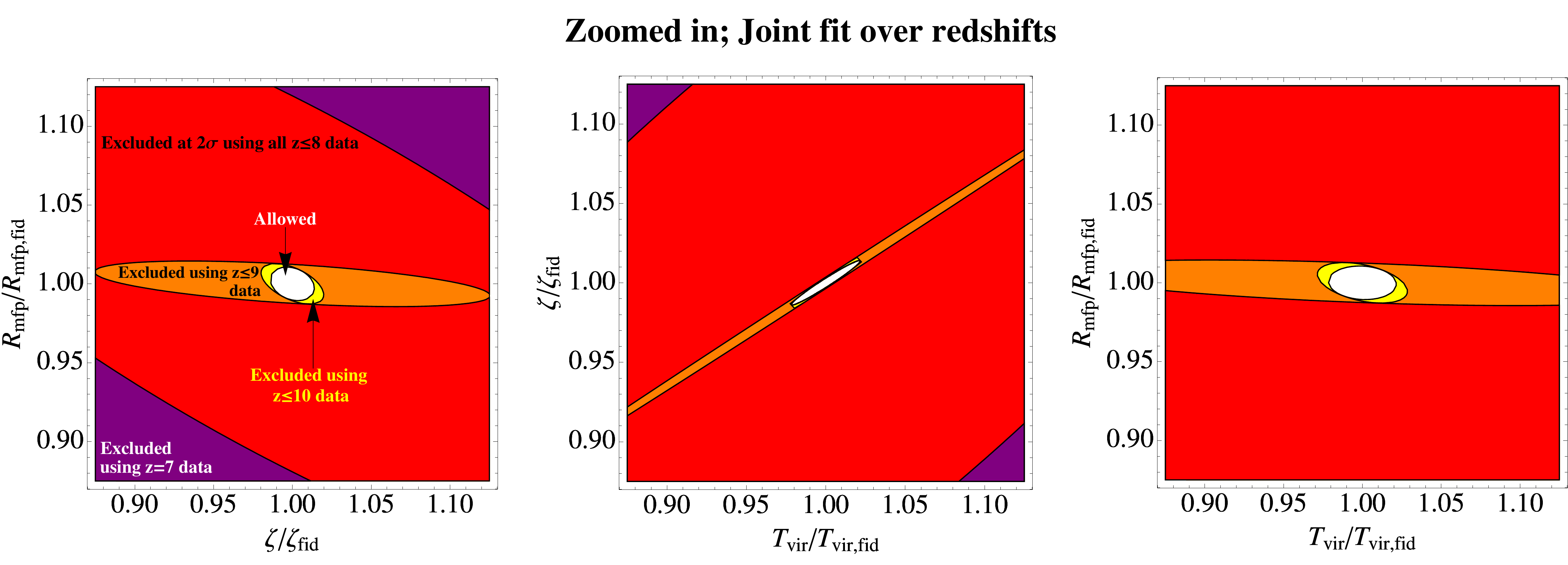}
\caption{Similar to Figure \ref{fig:MIDellipses}, but for the optimistic foreground model.  The top row shows the exclusion region from using $z=7$ data alone.  The middle and bottom rows show zoomed-in parameter space plots for the redshift-by-redshift and simultaneous fits respectively.  (Since $z=7$ is the lowest redshift in our model, the top panel is the same for both types of fit).  The constraints in this optimistic foreground scenario are seen to be better than those predicted for the moderate foreground model by about a factor of four.}
\label{fig:OPTellipses}
\end{figure*}

To see how degeneracies are broken by using information from multiple redshifts, imagine a thought experiment where one began with measurements at the lowest (least noisy) redshifts, and gradually added higher redshift information, one redshift at a time.  Figures \ref{fig:MIDellipses} and \ref{fig:OPTellipses} show the results for the moderate and optimistic foreground scenarios respectively.  (Here we omit the equivalent figure for the pessimistic model completely, because the results are again qualitatively similar to those for the moderate model.)  In each figure are $2\sigma$ constraints for pairs of parameters, having marginalized over the third parameter by assuming that the likelihood function is Gaussian (so that the covariance of the measured parameters is given by the inverse of the Fisher matrix).  One sees that as higher and higher redshifts are included, the principal directions of the exclusion ellipses change, reflecting the first degeneracy-breaking effect highlighted above, namely, the inclusion of different, more-complex and less-degenerate astrophysics at higher redshifts.  To see the second degeneracy-breaking effect, where per-redshift degeneracies are broken by joint redshift fits, we include in both figures the constraints that arise after combining results from redshift-by-redshift fits (shown as contours for each redshift), as well as the constraints from fitting multiple redshifts simultaneously (shown as cumulative exclusion regions).

For the moderate foreground scenario, we find that non-trivial constraints cannot be placed using $z=7$ data alone, hence the omission of a $z\le 7$ exclusion region from Figure \ref{fig:MIDellipses}.  However, we note that the constraints using $z \le 8$ data (red contours/exclusion regions) are substantially tighter in the bottom panel than in the top panel of the figure.  This means that the $z=7$ power spectrum can break degeneracies in a joint fit, even if the constraints from it alone are too degenerate to be useful.  A similar situation is seen to be true for a $z=10$ measurement, which is limited not just by degeneracy, but also by the higher thermal noise at lower frequencies.  Except for the $\zeta$-$T_\textrm{vir}$ parameter space, adding $z=10$ information in an independent fashion does not further tighten the constraints beyond those provided by $z \le 9$.  But again, when a joint fit (bottom panel of Figure \ref{fig:MIDellipses}) is performed, this information is useful even though it was noisy and degenerate on its own.  We caution, however, that this trend does \emph{not} persist beyond $z =10$, in that $z \ge 11$ measurements are so thermal-noise dominated that their inclusion has no effect on the final constraints.  Indeed, the ``allowed" regions in both Figures \ref{fig:MIDellipses} and \ref{fig:OPTellipses} include all redshifts, but are visually indistinguishable from ones calculated without $z\ge 11$ information.\footnote{We emphasize that in our analysis we have only considered the reionization epoch.  Thus, while we find that observations of power spectra at $z \ge 11$ do not add very much to measurements of reionization parameters like $T_\textrm{vir}$, $\zeta$, and $R_\textrm{mfp}$, they are expected to be extremely important for constraining X-ray physics prior to reionization, as discussed in \cite{mesinger_et_al_2013} and \cite{christian_and_loeb_2013}.}

\begin{figure*}[!ht]
\centering
\includegraphics[width=1.0\textwidth,trim=0cm 0cm 0cm 0cm,clip]{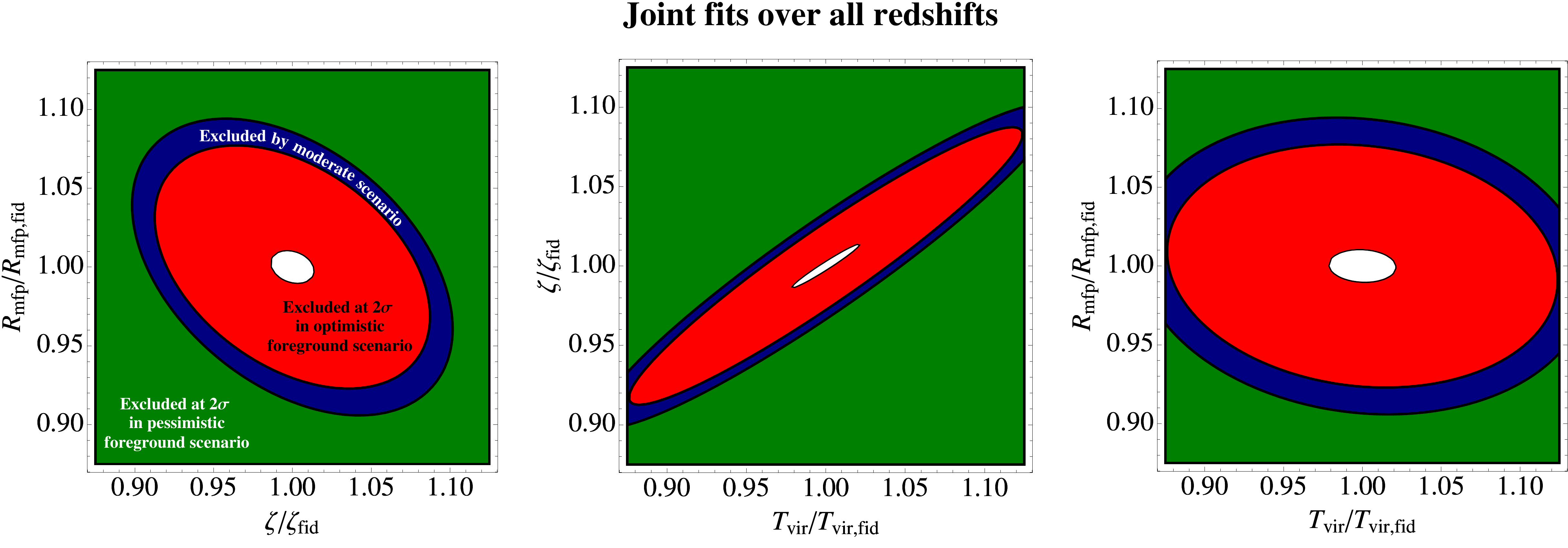}
\caption{A comparison between the $2\sigma$ exclusion regions for pessimistic (green), moderate (blue), and optimistic (red) foregrounds, assuming that power spectra at all measured redshifts are fit simultaneously.  Going from the pessimistic foreground model to the moderate model gives only marginal improvement; going from the moderate to the optimistic model reduces errors from the $5\%$ level to the $1\%$ level.}
\label{fig:OPTMIDPESSellipses}
\end{figure*}

Comparing the predictions for the moderate foreground model to those of the optimistic foreground model (Figure \ref{fig:OPTellipses}), several differences are immediately apparent.  Whereas the $z=7$ power spectrum alone could not place non-trivial parameter constraints in the moderate scenario, in the optimistic scenario it has considerable discriminating power, similar to what can be achieved by jointly fitting all $z \le 9$ data in the moderate model.  This improvement in the effectiveness of the $z=7$ measurement is due to an increased ability to access low and high $k$ modes, which breaks degeneracies.  With low and high $k$ modes measurable, each redshift alone is already reasonably non-degenerate, and the main benefit (as far as degeneracy-breaking is concerned) in going to higher $z$ is the opportunity to access new astrophysics with a slightly different set of degeneracies, rather than the opportunity to perform joint fits.  Indeed, we see from the middle and bottom panels of Figure \ref{fig:OPTellipses} that there are only minimal differences between the joint fit and the independent fits.  In contrast, with the moderate model in Figure \ref{fig:MIDellipses} we saw about a factor of four improvement in going from the latter to the former.

\begin{table}
\centering
\begin{tabular}{cccc}
Foreground Model & $\Delta T_\textrm{vir}/T_\textrm{vir,fid}$ & $\Delta \zeta$/$\zeta_\textrm{fid}$ & $\Delta R_\textrm{mfp}/R_\textrm{mfp,fid}$ \\
\hline
Moderate & $0.062 $ & $0.044$ & $0.039$ \\
Pessimistic & $0.071$ & $0.051$ & $0.047$ \\
Optimistic & $0.011$ & $0.0069$ & $0.0052$ \\
\end{tabular}
\caption{Reionization Parameter Errors ($1\sigma$) for HERA}
\label{finalErrors}
\end{table}

Figure \ref{fig:OPTMIDPESSellipses} compares the ultimate performance of HERA for the three foreground scenarios, using all measured redshifts in a joint fit.  (Note that our earlier emphasis on the differences between joint fits and independent fits was for pedagogical reasons only, since in practice there is no reason not to get the most out of one's data by performing a joint fit.)  We see that even with the most pessimistic foreground model, our three parameters can be constrained to the $5\%$ level.  The ability to combine partially-coherent baselines in the moderate model results only in a modest improvement, but being able to work within the wedge in the optimistic case can suppress errors to the $\sim 1 \%$ level.  The final results are given in Table \ref{finalErrors}.

In closing, we see that a next-generation like HERA should be capable of delivering excellent constraints on astrophysical parameters during the EoR.  These constraints will be particularly valuable, given that none of the parameters can be easily probed by other observations.  However, a few qualifications are in order.  First, the Fisher matrix analysis performed here provides an accurate forecast of the errors only if the true parameter values are somewhat close to our fiducial ones.  As an extreme example of how this could break down, suppose $T_\textrm{vir}$ were actually $1000\,\textrm{K}$, as illustrated in the middle row of Figure \ref{fig:pspecs}.  The result would be a high-redshift reionization scenario, one that would be difficult to probe to the precision demonstrated in this section, due to high thermal noise.  Secondly, one's ability to extract interesting astrophysical quantities from a measurement of the power spectrum is only as good as one's ability to model the power spectrum.  In this section, we assumed that \texttt{21cmFAST} is the ``true" model of reionization.  At the few-percent-level uncertainties given in Table \ref{finalErrors}, the measurement errors are better than or comparable to the scatter seen between different theoretical simulations \citep{zahn_et_al_2011}.  Thus, there will likely need to be much feedback between theory and observation to make sense of a power spectrum measurement with HERA-level precision.  Alternatively, given the small error bars seen here with a three-parameter model, it is likely that additional parameters can be added to one's power spectrum fits without sacrificing the ability to place constraints that are theoretically interesting.  We leave the possibility of including additional parameters (many of which have smaller, subtler effects on the $21\,\textrm{cm}$ power spectrum than the parameters examined here) for future work.

\section{Conclusions}
\label{sec:conclusions}

In order to
explore the potential range of constraints that will come from the proposed
next generation of 21~cm experiments (e.g. HERA and SKA), we used simple
models for instruments, foregrounds, and reionization histories to encompass
a broad range of possible scenarios.  For an instrument model, we used the 
$\sim 0.1~\rm{km}^2$ HERA concept array, and calculated power spectrum
sensitivities using the method of \cite{pober_et_al_2013a}.  To cover
uncertainties in the foregrounds, we used three principal models.  Both
our pessimistic and moderate model assumes foregrounds occupy the wedge-like
region in $k$-space observed by \cite{pober_et_al_2013b}, extending
$0.1~h{\rm Mpc}^{-1}$ past the analytic horizon limit. 
Thus, both cases are amenable to a strategy of foreground avoidance.
What makes our
pessimistic model pessimistic is the decision to combine partially redundant
baselines in an incoherent fashion, allowing one to completely sidestep
the systematics highlighted by \cite{hazelton_et_al_2013}. 
In the moderate model,
these baselines are allowed to be combined coherently.  Finally, in our
optimistic model, the size of the wedge is reduced to a region defined
by the FWHM of the primary beam.  Given the small field of view of the 
dishes used in the HERA concept array, this model is effectively equivalent
to one in which foreground removal techniques prove successful.  Lastly,
to cover the uncertainties in reionization history, we use \texttt{21cmFAST} to generate
power spectra for a wide range of uncertain parameters: the ionizing
efficiency $\zeta$, the minimum virial temperature of halos producing
ionizing photos, $T_{\rm vir}$, and the mean free path of ionizing photons
through the IGM $R_{\rm mfp}$.

Looking at predicted power spectrum measurements for these various scenarios
yields the following conclusions:
\begin{itemize}

\item Even with no development of analysis techniques beyond those used
in \cite{parsons_et_al_2013}, an experiment with
$\sim 0.1~\rm{km}^2$ of collecting area can yield very high significance
$\gtrsim 30\sigma$ detections of nearly 
any reionization power spectrum (cf. Figure
\ref{fig:inst-red-snr-matrix}).

\item Developing techniques that allow for the coherent addition of partially
redundant baselines can result in a small increase of additional
power spectrum sensitivity.  In this work, we find our moderate foreground
removal model to increase sensitivities by $\sim20\%$ over our most pessimistic
scenario.  Generally, we find that coherent combination of partially
redundant baselines reduces thermal noise errors by $\sim40\%$, so addressing
this issue will be somewhat more important for smaller arrays that have
not yet reached the sample variance dominated regime.

\item With the sensitivities achievable with our moderate foreground model,
the next generation of arrays will yield high significance detections of
the EoR power spectra, and provide detailed characterization of the power
spectrum shape over an order-of-magnitude in $k$ 
($k\sim0.1\mbox{--}1.0 h{\rm Mpc}^{-1}$). 
These sensitivity levels may even allow for direct imaging of the EoR
on these scales.

\item If successful, foreground removal algorithms can dramatically
boost the sensitivity of 21~cm measurements.  They are also the only
way to open up the largest scales of the power spectrum, which can
lead to new physical insight through observations of the generic ``knee" 
feature.

\item Although it will represent a major breakthrough for the 21~cm cosmology
community, a low to moderate ($\sim 5\mbox{--}10\sigma$) detection of the EoR power
spectrum may not be able to conclusively identify the redshift of 50\%
ionization.  One might expect otherwise, since the peak brightness of the power
spectrum occurs near this ionization fraction.  However, accounting for the steep
rise in $T_{\rm sys}$ at low frequencies, 
shows that the rise-and-fall
of the power spectrum versus redshift may not be conclusively measurable
without a higher significance measurement, such as those possible
with the HERA design.

\item Going beyond power spectrum measurements to astrophysical parameter constraints, lower redshifts observations are particularly prone to parameter degeneracies.  These can be partially broken by foreground removal from within the wedge region (allowing access to the lowest $k$ modes).  Alternatively, degeneracies can be broken by performing parameter fits over multiple redshifts simultaneously (which is equivalent to making use of information about the power spectrum's \emph{evolution}).  Higher redshifts ($z \ge 11$) are typically limited not by intrinsic degeneracies, but by high thermal noise (at least for a HERA-like array), and add relatively little to constraints on reionization.

\item Assuming a fiducial 21cmFAST reionization model, a HERA-like array will be capable of constraining reionization parameters to $\sim 1\%$ uncertainty if foreground removal within the wedge proves possible, and to $\sim 5\%$ otherwise.  The current generation of interferometers will struggle to provide precise constraints on reionization models; the sensitivity of a HERA-like array is necessary for this kind of science (for a quantitative comparison, see the appendix).

\end{itemize} 

From this analysis, it is clear that for 21~cm studies to deliver the 
first conclusive scientific constraints on the Epoch of Reionization, 
arrays much larger than those currently operational must be constructed.
Advancements in analysis techniques to keep
the EoR window free from contamination can contribute additional sensitivity,
but the most dramatic gains on the analysis front will come from
techniques that remove foreground emission and allow retrieval of modes
from inside the wedge.
This is not meant to disparage the wide range of foreground removal
techniques already in the literature; rather, the impetus is on adapting
these techniques for application to real data from the current and next
generation of 21~cm experiments.  The vast range of EoR science 
achievable under our optimistic, moderate, and even pessimistic foreground removal scenarios 
provides ample motivation for continuing these efforts.    

\acknowledgments{JCP is supported by an NSF Astronomy and Astrophysics 
Fellowship under award AST-1302774.  The authors wish to thank
our anonymous reviwer for their helpful feedback on the draft version of this
manuscript.  We also would like thank
the organizers of the 2013 CAASTRO science conference,
``Reionization in the Red Centre" where many of the ideas for this paper
were developed.}

\appendix
\label{app:others}
\section{Power Spectrum Sensitivities of Other 21~cm Experiments}
 
In this appendix, we compare the power spectrum sensitivities and 
EoR parameter constraints of several 21~cm experiments. In particular, we consider the current generation experiments of PAPER \citep{parsons_et_al_2010},
the MWA \citep{tingay_et_al_2013}, and 
LOFAR \citep{van_haarlem_et_al_2013}, as well as a concept array for Phase
1 of the SKA based on the SKA System Baseline Design document
(SKA-TEL-SKO-DD-001\footnote{http://www.skatelescope.org/wp-content/uploads/2012/07/SKA-TEL-SKO-DD-001-1\_BaselineDesign1.pdf}). 
The instrument designs are summarized
in Table \ref{tab:instruments}, and the principal results are presented
in Tables \ref{tab:signif} and \ref{tab:astrophys}, which show
the significance of the power spectrum measurements and constraints
on EoR astrophysical parameters, respectively.  
Both calculations assume the fiducial EoR history shown
in Figure \ref{fig:pspec_vanilla}.  The significances in 
Table \ref{tab:signif} assume only an 8~MHz band centered on the 50\%
ionization redshift of $z=9.5$.  The astrophysical constraints, however,
assume information is collected over a wider band from $z = 7\mbox{--}13$;
for instruments with smaller instantaneous bandwidths, the observing
times will need to be adjusted accordingly.\footnote{For the particulars
of our fiducial EoR model, a significant fraction of the information comes
from $z = 7\mbox{--}9$ (142\mbox{--}178~MHz) 
(see Figure \ref{fig:OPTellipses}), meaning that an experiment
like the MWA with an instantaneous bandwidth of 30~MHz could nearly produce
the results described here without a signficant correction
for observing time.  
Of course, this assumes that the redshift of reionization is known
a priori, and that the optimal band for constraints is actually the band
observed.}

\ctable[
caption=Properties of Other 21~cm Experiments, 
label=tab:instruments, 
pos=ht, 
mincapwidth=\textwidth,
]{c|p{.66in}p{.66in}p{.7in}p{3.3in}}{
\tnote[a]{Assumes each HBA sub-station is correlated independently.}
}{
\toprule
Instrument & Number of Elements & Element Size~$(\rm{m}^2)$ & Collecting Area~$(\rm{m}^2)$ & Configuration \\
\midrule
PAPER & 132 & 9 & 1188 & $11 \times 12$ sparse grid \\
MWA & 128 & 28 & 3584 & Dense 100~m core with $r^{-2}$ distribution beyond \\
LOFAR NL Core & 48\tmark[a] & 745 & 35,762 & Dense 2~km core \\
\midrule
HERA & 547 & 154 & 84,238 & Filled 200~m hexagon \\
SKA1 Low Core & 866 & 962 & 833,190 & Filled 270~m core with Gaussian distribution beyond \\
\bottomrule}

\begin{table}
\centering
\begin{tabular}{c|ccc} 
Instrument & Pessimistic & Moderate & Optimistic \\
\hline
PAPER & 1.17 & 2.02 & 4.82 \\
MWA & 0.60 & 2.46 & 6.40 \\
LOFAR NL Core & 1.35 & 2.76 & 17.37 \\
HERA & 32.09 & 38.20 & 133.15 \\
SKA1 Low Core & 10.01 & 35.95 & 218.27 \\
\end{tabular}
\caption{Power spectrum measurement signifiance (number of $\sigma$s) of other 21~cm experiments
for each of the three foreground removal models.}
\label{tab:signif}
\end{table}

\begin{table}
\centering
\begin{tabular}{c|ccc|ccc|ccc} 
\toprule
\multicolumn{1}{c}{} & \multicolumn{3}{c}{Pessimistic} & \multicolumn{3}{c}{Moderate} & \multicolumn{3}{c}{Optimistic} \\
Instrument & $\frac{\Delta T_{\rm vir}}{T_{\rm vir,fid}}$ & $\frac{\Delta \zeta}{\zeta_{\rm fid}}$ & $\frac{\Delta R_{\rm mfp}}{R_{\rm mfp,fid}}$ & $\frac{\Delta T_{\rm vir}}{T_{\rm vir,fid}}$ & $\frac{\Delta \zeta}{\zeta_{\rm fid}}$ & $\frac{\Delta R_{\rm mfp}}{R_{\rm mfp,fid}}$ & $\frac{\Delta T_{\rm vir}}{T_{\rm vir,fid}}$ & $\frac{\Delta \zeta}{\zeta_{\rm fid}}$ & $\frac{\Delta R_{\rm mfp}}{R_{\rm mfp,fid}}$ \\
\midrule
PAPER & 1.444 & 1.168 & 1.507 & 1.260 & 1.013 & 1.294 & 0.272 & 0.179 & 0.140 \\
MWA & 4.419 & 3.479 & 4.555 & 0.757 & 0.568 & 0.731 & 0.231 & 0.152 & 0.119 \\
LOFAR & 1.538 & 1.251 & 1.515 & 0.719 & 0.565 & 0.675 & 0.069 & 0.046 & 0.039 \\
HERA & 0.072 & 0.051 & 0.047 & 0.062 & 0.044 & 0.039 & 0.011 & 0.007 & 0.005 \\
SKA1 & 0.235 & 0.169 & 0.179 & 0.076 & 0.054 & 0.044 & 0.009 & 0.006 & 0.004 \\
\end{tabular}
\caption{Fractional errors on the reionization parameters achieveable with each instrument under the three foreground removal models,
assuming all redshifts are analyzed jointly.}
\label{tab:astrophys}
\end{table}

In order to compute the constraints achievable with other experiments,
we apply the sensitivity calculation described in \S\ref{sec:sense_calc}
to each of the five instruments under study.
We note that this sensitivity calculation assumes a drift-scanning
observing mode, with the limit of coherent sampling set by the
size of the element primary beam.  The MWA, LOFAR, and likely the SKA 
all have the capability of conducting a tracked scan to increase the
coherent integration on a single patch of sky.  Similarly, tracking can
be used to move to declinations away from zenith if sample variance
becomes the dominant source of error.  A full study of the benefits
of tracking versus draft scanning for power spectrum measurements is 
beyond the scope of this present work; rather, we assume all instruments
operate in a drift-scanning mode for the clearest comparison
with the fiducial results calculated for the HERA experiment.
We therefore also assume that each telescope observes for the fiducial
6~hours~per~day for 180~days (1080~hours).  
Finally, we also assume that 
each array has a receiver temperature of 100~K.
We discuss the important features of each instrument and the resultant
constraints in turn; see the main text for a discussion of the HERA
experiment.

\begin{enumerate}

\item \textbf{PAPER}: Our fiducial PAPER instrument is an $11\times12$
grid of PAPER dipoles modelled after the maximum redundancy arrays
presented in \cite{parsons_et_al_2012a}.  In this configuration,
the $3\times3~\rm{m}$ dipoles are spaced in 12 north-south columns
separated by 16~m; within a column, the dipoles are spaced 4~m apart. 
In both our pessimistic and moderate scenarios, PAPER yields a
non-detection of the fiducial 21~cm power spectrum.  In the optimistic
scenario, the array could yield a significant detection; however,
the poor PSF of the maximum redundancy array is expected to present
challenges to any foreground-removal strategy that would allow
recovery from information inside the wedge \citep{parsons_et_al_2013}.
Therefore, achieving the results of the optimistic scenario will be especially
difficult for the PAPER experiment.

\item \textbf{MWA}: Our model MWA array uses the 128 antenna positions
presented in \cite{tingay_et_al_2013}.  Despite having nearly three times
the collecting area of the PAPER array, we find the MWA yields a 
less significant detection in the pessimistic scenario.  
Poor sensitivity when partially redundant samples are combined
incoherently is to be expected for the MWA.  
The pseudo-random configuration of the array produces
essentially no instantaneously redundant samples, and so all
redundancy comes from partial coherence.  Therefore, one might expect
the MWA to under-perform compared to the highly redundant PAPER array in
this scenario. In the moderate and optimistic scenarios
where partial redundancy yields sensitivity boosts
the MWA outperforms the PAPER array.

\item \textbf{LOFAR}: To model the LOFAR array, we use the antenna
positions presented in \cite{van_haarlem_et_al_2013}.  For the purposes
of EoR power spectrum studies, we focus on the Netherlands core
of the instrument, since baselines much longer than a few km contribute
very little sensitivity.  We also assume that LOFAR is operated in a mode
where each sub-station of the HBA is correlated separately to increase
the number of short baselines.  However, the resultant sensitivities
still show that LOFAR suffers from a lack of short baselines.  Despite
having a collecting area $\gtrsim10$ times larger than PAPER and the MWA,
LOFAR still yields a non-detection of the EoR power spectrum
in the pessimistic and moderate foreground removal scenarios.  Only
in the optimistic scenario where longer baselines contribute
to the power spectrum measurements does LOFAR's collecting area
result in a high-significance measurement.  Preliminary results from the
LOFAR experiment show significant progress in subtracting
foregrounds to access modes inside the wedge 
\citep{chapman_et_al_2012,yatawatta_et_al_2013}.

\item \textbf{SKA1-Low}: We model our SKA-Low Phase 1 instrument
after the design parameters set out in the SKA System Baseline
Design document, although the final design of the SKA is still subject
to change.  This document specifies that the array will consist of 911
35~m stations, with 866 stations in a core with a Gaussian distribution
versus radius.  This distribution is normalized to have 650 stations
within a radius of 1~km.  This density in fact yields a completely
filled aperture out to $\sim300~\rm{m}$, which we model as a close
packed hexagon.
This core gives the design some degree of instantaneous redundancy,
a configuration
that is still being explored for the final design of the instrument.
We do not consider the 45 outriggers in our power
spectrum sensitivity.  Much like the case with PAPER and the MWA, the
lower instantaneous redundancy of the SKA concept array results in a poorer
performance than the highly redundant HERA array in the pessimistic
scenario.  However, in the moderate and optimistic scenarios this
SKA concept design yields high sensitivity measurements, although
not as high as might be expected from collecting area alone.  This
fact is once again due to the relatively small number of short spacings
compared to the HERA array, resulting in similar performances for the two
configurations in the moderate scenario.  As with LOFAR, the SKA design
shines in the optimistic scenario, producing a very high SNR measurement.

\end{enumerate}

\begin{figure*}[t]
\centering
\includegraphics[width=1.0\textwidth,trim=2cm 0cm 2cm 1.25cm,clip]{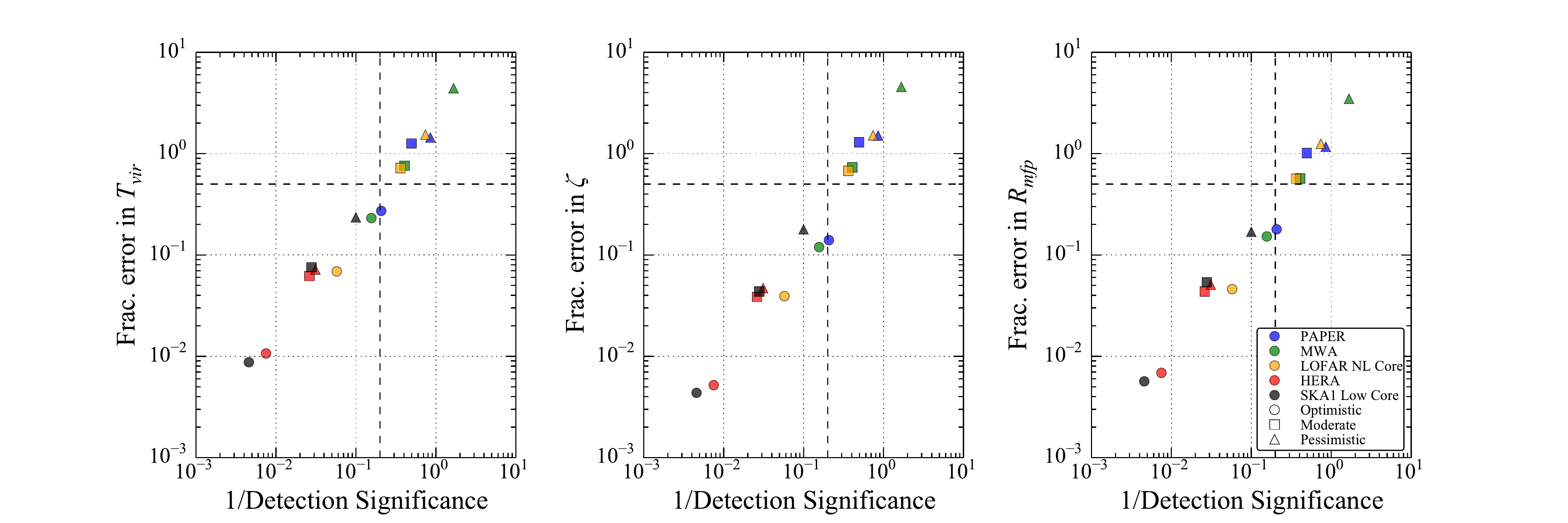}
\caption{Fractional errors in astrophysical parameters shown as a function of $(\textrm{detection significance})^{-1}$.  Different instruments are shown in different colors (PAPER in blue; MWA in green; 
LOFAR in yellow; HERA in red; SKA in black), and different foreground scenarios are shown using
different shapes (optimistic as circles; moderate as squares; pessimistic as triangles).  The vertical
dashed line delineates a $5\sigma$ detection of the power spectrum, while the horizontal
dashed line delineates a parameter error of $50\%$.  The tight correlations shown here suggest that the significance of a power spectrum detection can be used as a proxy for an instrument's ability to constrain
astrophysical parameters.}
\label{fig:sigmaParamCorrAllz}
\end{figure*}
\vspace{.2in}
In all cases, we find that the fractional errors on the reionization parameters
(Table \ref{tab:astrophys}) scale very closely with the overall
significance of the power spectrum measurement (Table \ref{tab:signif}).
This is shown in Figure \ref{fig:sigmaParamCorrAllz}, where we plot the fractional
errors on the reionization parameters against the reciprocal of the power spectrum
detection significance.  These two quantities are seen to be directly proportional
to an excellent approximation, regardless of foreground scenario.\footnote{We note that this is true only when all redshifts are analyzed jointly, 
where the errors are driven mainly by thermal noise.  If the errors are instead
dominated by parameter degeneracies (as is the case, for example, when
only one redshift slice is measured), the tight linear correlation breaks down.}  
Therefore, while the power spectrum sensitivity of an array can 
be a strong function of an instrument's configuration,
the resultant astrophysical constraints are fairly generalizable once
the instrument sensitivity is known.  This is strongly suggestive
that the results in the main body of the paper can be easily extended
to other instruments.  Also noteworthy is the fact that current-generation
instruments encroach on the lower-left regions (high detection significance;
 small parameter errors) of the plots in Figure \ref{fig:sigmaParamCorrAllz}
 only for the optimistic foreground model.  In contrast, the next-generation
 instruments (HERA and SKA) are clearly capable of delivering excellent
 scientific results even in the most pessimistic foreground scenario.
%

\bibliographystyle{apj}
\bibliography{nextgen}{}

\end{document}